\documentclass[journal=aelccp,manuscript=article]{achemso}

\usepackage[version=3]{mhchem} 
\usepackage[T1]{fontenc} 

\usepackage{bm}
\usepackage{xr}
\graphicspath{{./}}
\usepackage[usenames]{color}
\usepackage{hyperref}
\pdfoutput=1

\newcommand{\MAPI}{CH$_3$NH$_3$PbI$_3$}
\newcommand{\MAPBr}{CH$_3$NH$_3$PbBr$_3$}
\newcommand{\MAPCl}{CH$_3$NH$_3$PbCl$_3$}
\newcommand{\MA}{CH$_3$NH$_3^+$}

\newcommand{\CsPCl}{CsPbCl$_3$}

	\author{Alexander N. Beecher}
	\altaffiliation{Contributed equally to this work}

	\author{Octavi E. Semonin}
	\altaffiliation{Contributed equally to this work}
	\affiliation{Department of Chemistry, Columbia University, New York, NY 10027, USA}

	\author{Jonathan M. Skelton}
	\author{Jarvist M. Frost}
	\affiliation{Department of Chemistry, University of Bath, Claverton Down, Bath BA2 7AY, UK}

	\author{Maxwell W. Terban}
	\author{Haowei Zhai}
	\affiliation{Department of Applied Physics and Applied Mathematics, Columbia University, New York, NY 10027, USA}
	
	\author{Ahmet Alatas}
	\affiliation{Advanced Photon Source, Argonne National Laboratory, Argonne, IL, 60439, USA}

	\author{Jonathan S. Owen}
	\affiliation{Department of Chemistry, Columbia University, New York, NY 10027, USA}

	\author{Aron Walsh}
	\affiliation{Department of Chemistry, University of Bath, Claverton Down, Bath BA2 7AY, UK}

	\author{Simon J. L. Billinge}
	\email{sb2896@columbia.edu}
	\affiliation{Department of Applied Physics and Applied Mathematics, Columbia University, New York, NY 10027, USA}
	\alsoaffiliation{Condensed Matter Physics and Materials Science Department, Brookhaven National Laboratory, Upton, NY 11973, USA}

	\title{Direct Observation of Dynamic Symmetry Breaking above Room Temperature in Methylammonium Lead Iodide Perovskite}

\begin{document}

\begin{abstract}
Lead halide perovskites such as methylammonium lead triiodide (\MAPI) have
outstanding optical and electronic properties for photovoltaic
applications, yet a full understanding of how
this solution processable material works so well is currently missing. Previous 
research has revealed that \MAPI\ possesses multiple forms of static disorder regardless of preparation method, which is surprising in light of its excellent performance.
Using high energy resolution inelastic X-ray (HERIX) scattering,
we measure phonon dispersions in \MAPI\ and
find direct evidence for another form of disorder in \emph{single} crystals: large
amplitude anharmonic zone-edge rotational instabilities of the PbI$_6$
octahedra that persist to room temperature and above,
left over from structural phase transitions that take place tens to hundreds of degrees below.
Phonon calculations show that
the orientations of the methylammonium (\MA) couple strongly and cooperatively to these modes.
The result is
a non-centrosymmetric, instantaneous local structure, which we observe in atomic pair distribution function (PDF) measurements.
This local symmetry breaking is unobservable by
Bragg diffraction, but can
explain key material properties such as the structural phase sequence, ultra low thermal
transport, and large minority charge carrier lifetimes despite
moderate carrier mobility.
\end{abstract}

\begin{center}
	\includegraphics{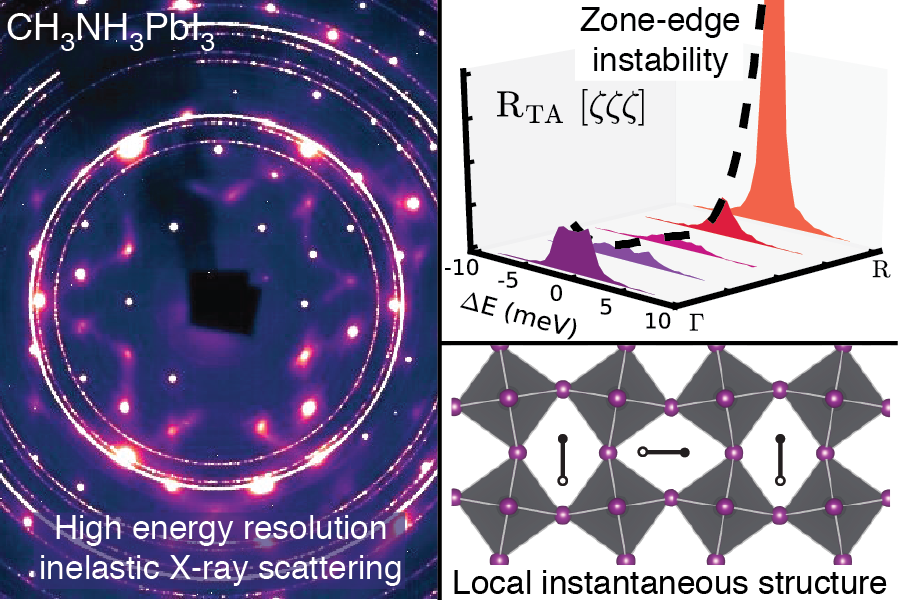}
\end{center}

Structural imperfections normally reduce the photovoltaic action of a material
by reducing the carrier mobilities and providing non-radiative recombination
pathways for the photo-generated carriers. \MAPI\ exhibits significant
nanocrystallinity\cite{Choi:2014ui}, defects~\cite{AronWalsh:2015cq} and
dynamic disorder~\cite{Poglitsch:1987ig,Yaffe:2016wt}; characteristics not
normally associated with high efficiency photovoltaic devices~\cite{Stranks:2015ef}.
The puzzle with hybrid halide perovskites is how such a defective solution
processed material can have efficiencies rivaling those of high quality
crystalline semiconductors~\cite{Green:2016fj}.
Two kinds of framework structural instabilities are
expected to be present in
perovskites: octahedral tilting~\cite{Glazer:1975er}, which is associated
with antiferroelectricity; and cation off-centering,
which can sometimes yield a ferroelectrically active distortion~\cite{Benedek:2013jl}.
Polarity in the material can affect the optical and electrical
properties~\cite{Frost:2014dm,Ma:2015he,Zhu:2015eh,liu_ferroelectric_2015}, but the presence of persistent
polarity in these materials has not been
established and continues to be disputed~\cite{Stoumpos:2013wq,Baikie:2015ju,stroppa_ferroelectric_2015,beilsten-edmands_non-ferroelectric_2015}.  In \CsPCl, octahedral rotational instabilities have been directly observed by inelastic neutron scattering~\cite{Fujii:1974di} and more indirectly for \MAPBr\ and \MAPCl~\cite{Swainson:2003hy,Chi:2005if,Swainson:2015ez}.
However, in \MAPI, while the disorder of the organic cation has been extensively
investigated~\cite{wasylishen_cation_1985, Poglitsch:1987ig, Stoumpos:2013wq,Leguy:2015hq,Chen:2015jp}, only calculations combined with indirect measurements have predicted octahedral rotational instabilities in the cubic phase~\cite{Quarti:2014wp,Brivio:2015dq,quarti_structural_2016}. In this work, we directly observe the lattice dynamics related to these framework distortions and connect them to the physical properties of \MAPI.

Inelastic scattering is a standard method to quantify these kinds of dynamics.
However, inelastic neutron scattering experiments (INS) require large single crystals
and are additionally challenging on hybrid
materials such as \MAPI\ due to the strong incoherent scattering of hydrogen.
We circumvent these issues by using X-ray based HERIX, which has a larger
scattering cross section and sensitivity, and better selectivity for
motion of the inorganic framework.

\begin{figure}
		\includegraphics{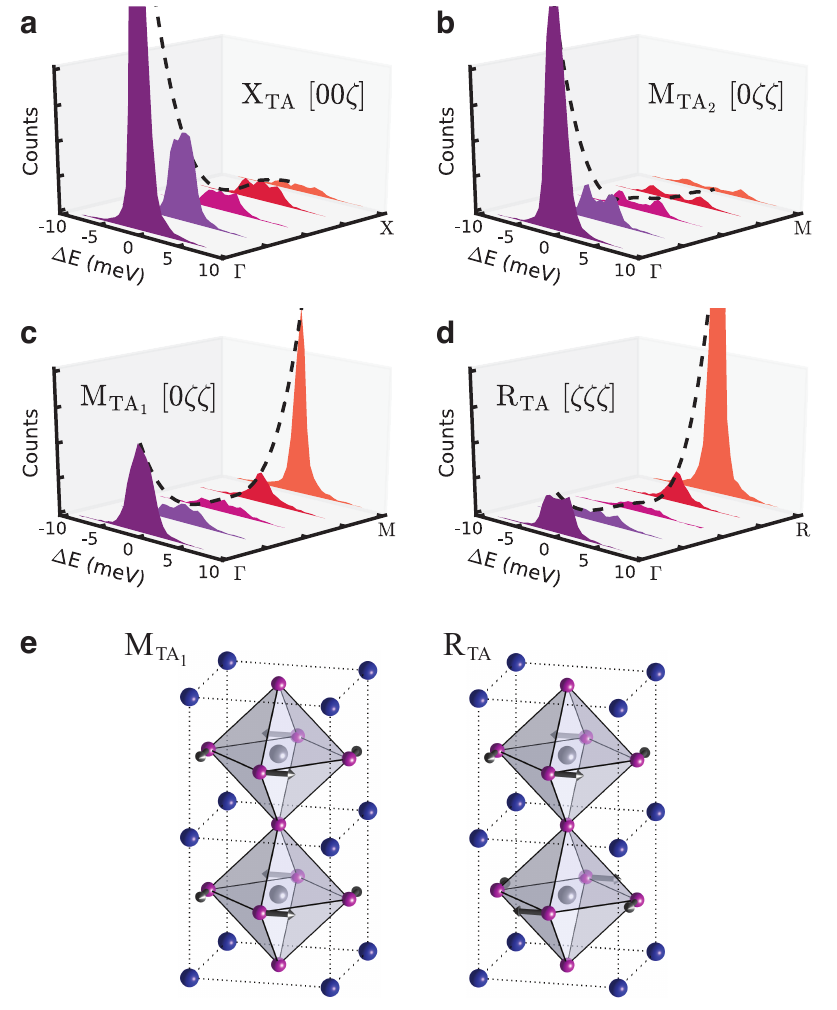}
		\caption{Inelastic scattering spectra plotted from the Brillouin zone-center to the zone-edge. \textbf{(a-d)} $M_{TA_2}$ and $X_{TA}$ illustrate what a typical dependence looks like. The dashed curve is only to guide the eye. In contrast, the $M_{TA_1}$ and $R_{TA}$ at the zone-edge have magnitude larger even than the Bragg tail.
		\textbf{(e)} Sketch of the motion of the observed anharmonic modes with the idealised A-site cation (\MA) position represented in blue, the B-site cation (Pb$^{2+}$) in grey at the center of the octahedra, and the X-site anion (I$^-$) in purple. Animations of these modes, and several others, are online (\textcolor{blue}{\href{https://figshare.com/s/97a6cbc033b17aa83a18}{Figshare}}). }
		\label{fig:3D}
	\end{figure}

Measurements were performed on high quality single crystals of \MAPI\ at the Advanced Photon Source at the Argonne National Laboratory.
Due to the large absorption cross-section of lead and iodine, crystals were polished to about 100 $\mu$m and mounted on a copper post (Fig.~S\ref{fig:xtal}A). This polishing and mounting preserves the high quality single crystal, as shown by single crystal X-ray diffraction (Fig.~S\ref{fig:xtal}B).
Measurements were performed at 350~K, in the cubic ($Pm3m$) phase, which gave a reasonable phonon intensity and removed difficulties associated with crystal twinning. Transverse acoustic (TA) and longitudinal acoustic (LA) phonon energies were measured along the three high-symmetry directions of the Brillouin zone, and one transverse optical (TO) branch was also measured. Representative plots of the raw spectra are shown in Fig.~\ref{fig:3D}.

The transverse acoustic branch in the $[00\zeta]$ direction ($X$) is shown in Fig.~\ref{fig:3D}a. Close to the zone-center ($\zeta=0$), the strongest signal is a resolution-limited elastic line coming from the tail of the nearby Bragg peak. Moving across the zone to the zone-edge, the elastic Bragg tail quickly dies off in intensity (indicated by the dashed line) and inelastic shoulders coming from the low energy acoustic modes become well resolved peaks at higher energy transfer ($\hbar\omega$). These peaks lie symmetrically on the energy-loss (Stokes) and energy-gain (anti-Stokes) side of $\hbar\omega=0$, corresponding to phonon creation and annihilation, respectively.
Approaching the zone-edge, the overall intensity of the spectrum is strongly
suppressed due to the lower phonon occupancy of high energy modes, amongst other effects. This is the expected behavior for the HERIX spectra of well defined acoustic phonons: dispersing to higher energy and decreasing in intensity with increasing $\zeta$.

However, this behavior is not observed in two of the phonon modes. In the $[0\zeta\zeta]$ direction the two transverse modes are non-degenerate and we refer to them as $M_{TA_1}$ and $M_{TA_2}$ (under cubic symmetry, the two transverse acoustic phonons are degenerate in both the $[00\zeta]$ and $[\zeta\zeta\zeta]$ directions). The intensity of the $M_{TA_2}$ branch behaves normally, similar to the transverse $[00\zeta]$ mode (Fig.~\ref{fig:3D}b), but the $M_{TA_1}$ mode behaves very differently as the zone is crossed from center to edge.
Now, approaching the zone-edge, a strikingly large broad central peak emerges at $\zeta=0.4$ and becomes narrower and very intense at $\zeta=0.5$ (Fig.~\ref{fig:3D}c).
The large signal intensity results from the low energy, and therefore high phonon occupation, of these modes.
Even more dramatic behavior is seen in the response of the $R_{TA}$ mode in the $[\zeta\zeta\zeta]$ direction, with a resolution-limited $\hbar\omega=0$ peak at the $R$-point of intensity twelve times the corresponding peak at $\zeta = 0.1$ (Fig.~\ref{fig:3D}d). These two modes are the most important features of our observed lattice dynamics and correspond to rotation of the octahedra along the principal cubic axes, with neighboring octahedra along the rotation axis either rotating together ($M_{TA_1}$) or opposite (in-phase and out-of-phase tilting respectively) to each other ($R_{TA}$). This motion is illustrated in Fig.~\ref{fig:3D}e~\cite{Fujii:1974di}.

\begin{figure}
		\includegraphics[width=80mm]{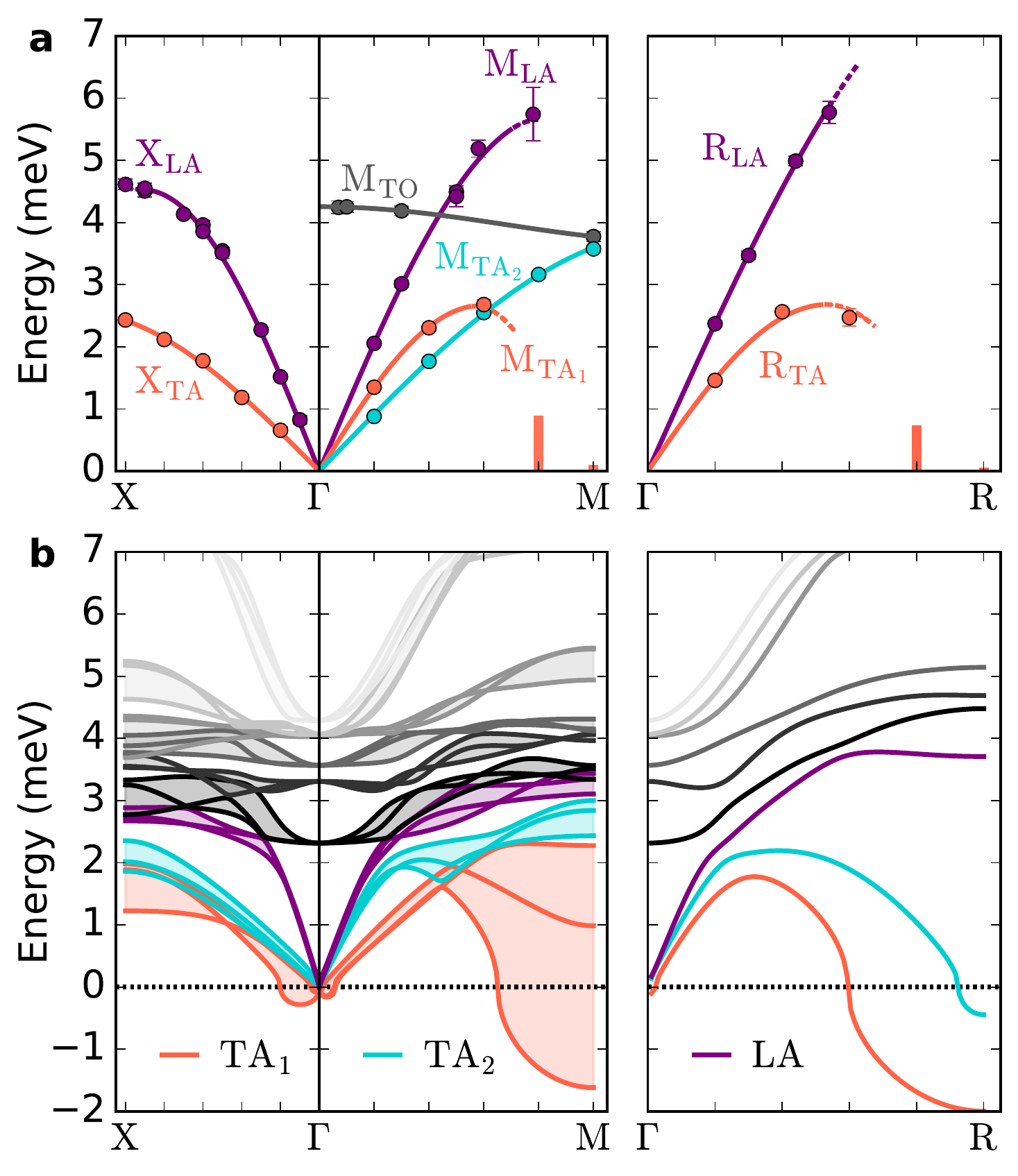}
		\caption{Measured and calculated dispersion curves. \textbf{(a)} The
        $X$, $M$, and $R$ longitudinal acoustic (LA) and transverse acoustic
        (TA) modes, and one transverse optical (TO) branch measured by HERIX
        are plotted. The $1\sigma$ uncertainty in $\zeta$ and phonon energy (by
        fitting) are generally less than the size of the marker. The lines
        (sine fits) are only to guide the eye. At the zone-edge of $M$ and $R$
        the phonon energy becomes small and overdamped, so we fit to
        a Lorentzian peak convoluted with the resolution function and centered
        at $\hbar\omega=0$~meV, and plot bars with height equal to the half
        width at half maximum of the Lorentzian. \textbf{(b)} Calculated phonon
        dispersion curves become imaginary along $M_{TA_1}$ and $R_{TA}$ as expected for a soft mode.}
		\label{fig:dispersion}
	\end{figure}

We have extracted phonon dispersions for the seven non-degenerate acoustic branches and one transverse optic branch (Fig.~\ref{fig:dispersion}).
Details of the fitting are presented in the Methods and Fig.~S\ref{fig:fitting}.
From the initial slope of the acoustic phonons, we extract elastic constants
and the bulk modulus ($K=13 \pm 2$ GPa), which implies that \MAPI\ has
a softness similar to wood~\cite{Gindl:2002il} (Table~S\ref{tab:elastics}). The width ($\Gamma$) of the Lorentz oscillator lineshape used to fit the modes is related to the phonon lifetime, $\tau = h/\Gamma$.
This analysis yields phonon lifetimes between 0.8 and 20 ps. Interestingly,
these lifetimes are comparable to the residence time of \MA\ in different
preferred orientations as measured by quasi-elastic neutron scattering (QENS) at room temperature~\cite{Leguy:2015hq}.
Combined with the anharmonic phonon modes, these short phonon lifetimes explain the ultra-low thermal conductivity~\cite{Pisoni:2014jy}, as in related lead chalcogenides~\cite{Delaire:2011gr, Li:2015gf}.

We now turn to first-principles lattice dynamics calculations of the phonon spectrum.
The calculated and measured dispersions along the three directions are in good agreement, as evident in  Fig.~\ref{fig:dispersion}.
The strong softening of the $M_{TA_1}$ and $R_{TA}$ modes is predicted by the athermal harmonic calculations,
where the mode frequencies become imaginary at $\zeta = 0.3$, close to where the onset of mode softening is observed in experiment.

The spread (shaded regions in Fig.~\ref{fig:dispersion}b) in the calculated dispersion curves arise
from the anisotropy of the molecule,
which breaks the degeneracy of the high-symmetry points in the Brillouin zone.
This shows very different restoring forces and mode energies calculated depending on the orientation of the ion
in the cage, indicating a strong coupling of the \MA\ dynamics to the cubo-octahedral cage in which it resides.
The spread, and therefore the coupling, is largest at the zone edge (Fig.~\ref{fig:dispersion}b) where the modes
soften to zero frequency and are found to be highly anharmonic.

\begin{figure}		
		\includegraphics{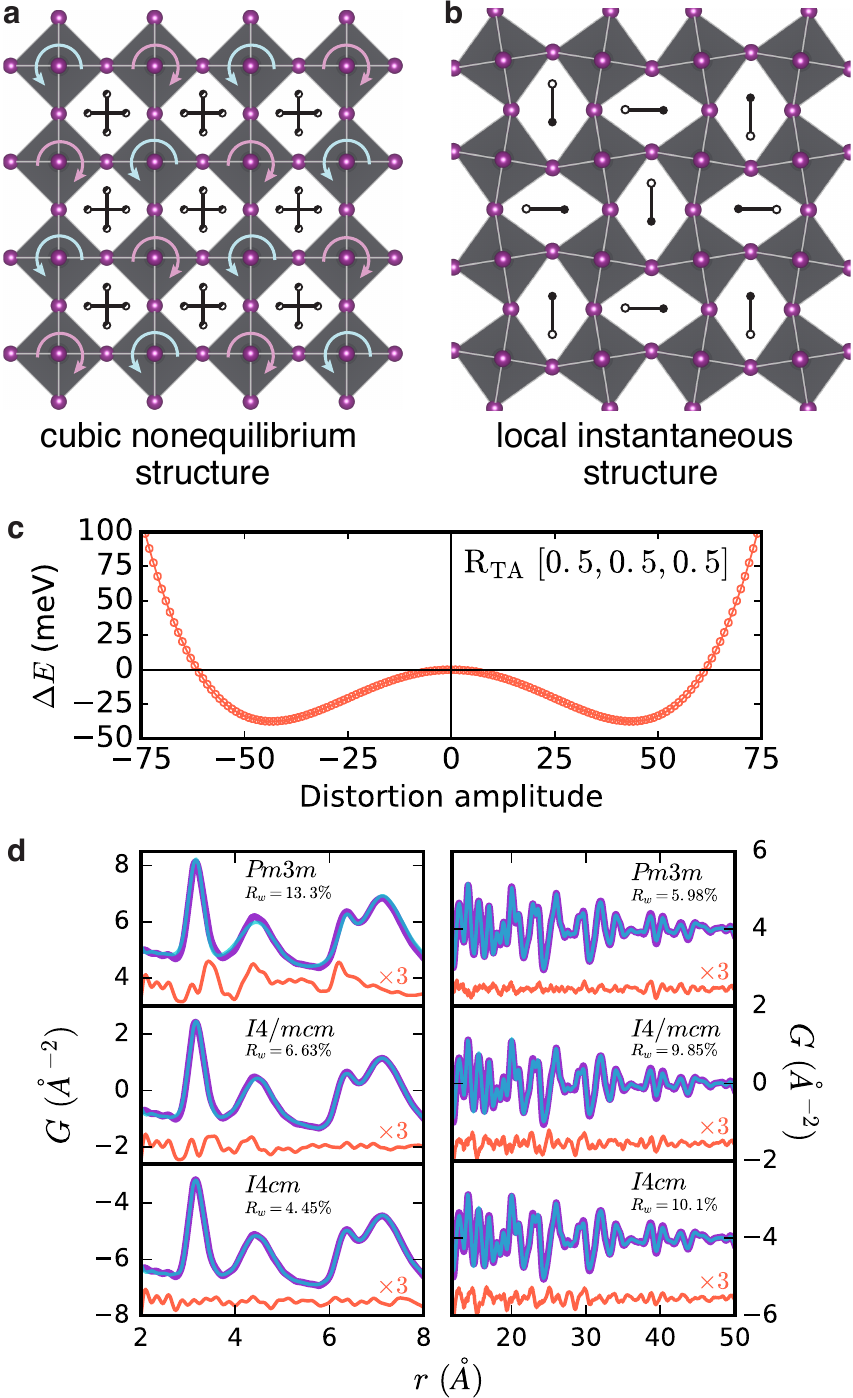}
		
		\caption{Local symmetry breaking in \MAPI\ at 350~K. \textbf{(a,b)} Distortions from cubic symmetry generate anisotropic cavities and couple to motion of the \MA\ ion, which we represent as off-centered and oriented along the long-axis of the cavity.
		\textbf{(c)} DFT-based lattice dynamic calculations show that the energy minimum at the $R$-point at 350~K is displaced in a double-well potential that causes local symmetry breaking.
		\textbf{(d)} Comparison of the experimental PDF (purple) to cubic ($Pm3m$), centrosymmetric ($I4/mcm$), and non-centrosymmetric ($I4cm$) tetragonal models (blue) show a superior fit for the low-symmetry models at low-$r$ (2-8 $\text{\AA}$). However, the models perform oppositely at high-$r$ with the high-symmetry cubic structure giving the best agreement to the data in the 12-50 $\text{\AA}$ region. The residuals (orange) are scaled $\times3$ for clarity.}
		\label{fig:pdf}
	\end{figure}

Considered alongside QENS measurements~\cite{Leguy:2015hq} that indicate a hopping rotational dynamics of the \MA, we conclude that the rotational motions of the cation and the cage dynamics are cooperative with important implications discussed below.
Octahedral tilting varies the shape of the perovskite A-site cavity in which the \MA\ ions reside.
As the cavity distorts, it elongates along one direction and is shortened in the perpendicular direction (Fig.~\ref{fig:pdf}a,b).
The above analysis suggests that the \MA\ ions stay aligned within locally-distorted A sites,
and only fluctuate between different local minima of the distorted cavity, cooperatively, on picosecond timescales.
On a timescale important for charge carriers (5~fs carrier scattering time~\cite{Karakus:2015de})
and at solar-relevant temperatures, the crystal structure is effectively frozen in local metastable symmetry-broken domains.

The computed potential landscapes of the anharmonic modes are indeed displaced minima
of shallow double-well potentials (Fig.~\ref{fig:pdf}c and Fig.~S\ref{fig:potentials}), consistent with our observation of a central peak in the inelastic spectra.
This dynamic symmetry breaking (see animations in the Supporting Information and \textcolor{blue}{\href{https://figshare.com/s/97a6cbc033b17aa83a18}{Figshare}})
is also evident in room temperature \textit{ab initio} molecular dynamics simulations
where persistent octahedral tilting away from the high-symmetry cubic orientation is observed~\cite{Frost:2016kl}.

Further evidence for this hypothesis is provided by X-ray atomic pair-distribution function (PDF) analysis of these materials. We discover that the low-$r$ region (where $r$ is the inter-atomic distance)
of the PDF is better fit by low symmetry tetragonal models than the cubic one, even at 350~K (Fig.~\ref{fig:pdf}d), indicating that the local structure is best described by tilted octahedra. The PDF refinement is further improved at low-$r$ when Pb is allowed to displace (0.041~\text{\AA}) from the high-symmetry position (Fig.~\ref{fig:pdf}d, bottom row).
When we zoom out to the high-$r$ region, we discover that a cubic model gives a better fit (Fig.~\ref{fig:pdf}d, second column), as expected due to averaging over dynamic differently-oriented symmetry-broken domains.  By performing PDF fits across different refinement ranges (Fig.~S\ref{fig:refranges}), we estimate the domain size to be 1-3~nm in diameter. 
Given the off-centering of the Pb and the methylammonium~\cite{Weller:2015jo,Ren:2016cw}, these domains may be polar.

The anharmonic modes indicate an incipient phase transition to the symmetry broken phases that emerge at lower temperatures, but with diffusive (order-disorder) dynamics persisting many tens to hundreds of Kelvin above the transition temperatures.
This is further supported by the intensity of the $X_{TA}$, $M_{TA_1}$, and $R_{TA}$ zone-edge peaks as a function of temperature.
	\begin{figure}
		\includegraphics[width=85mm]{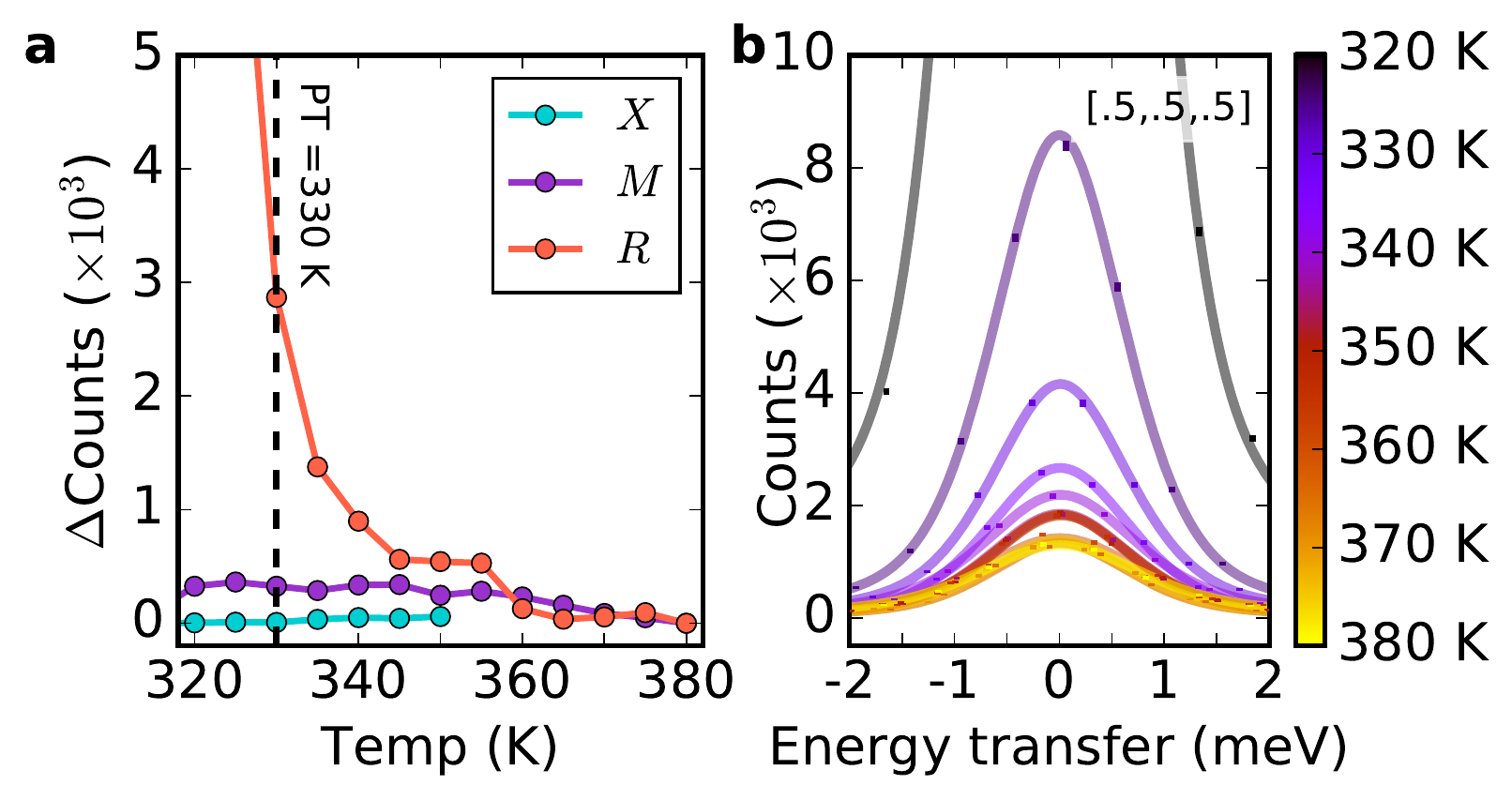}
		\caption{Temperature dependence of inelastic scattering spectra. \textbf{(a,b)} Intensity as a function of temperature is measured at the $X$-, $M$-, and $R$-points ($\Delta$Counts is equal to intensity of the measured mode at a given temperature less the intensity of the least intense peak in the temperature series). Scattering at the $X$-point has little temperature dependence while scattering at the $M$-point slightly increases in intensity with decreasing temperature. In contrast, scattering at the $R$-point responds dramatically, increasing sharply upon approaching the phase transition temperature due to the emergence of a Bragg peak in the tetragonal phase. This response indicates that the phase transition is driven by condensation of the $R_{TA}$ mode at 330 K.}
		\label{fig:nature}
	\end{figure}
As shown in Fig.~\ref{fig:nature}, there is no change in the intensity of the zone-edge $X_{TA}$, but the intensity of the zone-edge $R_{TA}$ intensity diverges sharply through the 330~K phase transition as a Bragg peak of the new tetragonal ($I4/mcm$) phase grows in. 
There is little change in the intensity of the zone-edge $M_{TA_1}$ at this temperature, suggesting that the $M_{TA_1}$ mode is related to the lower temperature phase transition at 160~K, reversing the sequence observed in \CsPCl~\cite{Fujii:1974di}. Although it was not possible to track this peak to 160~K due to the difficulty of aligning a sample with changing lattice constants, this finding is supported by crystallography~\cite{Baikie:2015ju} where a Bragg peak is observed in the low-temperature phase at the pseudo-cubic $M$-point of the parent structure.

We now explore implications of the discovery of soft anharmonic motion in \MAPI.
First, the soft anharmonic modes provide a large bath of acoustic phonons that are available for scattering and thermalising carriers. The population of low-energy phonons may explain the finding that electrical transport in these materials is phonon-limited~\cite{Karakus:2015de}.
In \MAPI, the electronic band extrema are in the vicinity of the $R$ point,
though made slightly indirect by the Rashba interaction\cite{Brivio2014.PRB}.
As well as local intravalley scattering (by acoustic and optical phonons at $\Gamma$), there is the possibility of intervalley scattering from phonon modes at the Brillouin zone boundary.
These are low in energy with a large occupancy at room temperature, suggesting that intervalley scattering
may be significant.
In the material GaP\cite{Kocsis1975}, where the band
extrema have multiple valleys, intervalley scattering dominates mobility above
200~K.

Second, these anharmonic modes point to a general model of the structural phase sequence in lead halide perovskites. The cubic-to-tetragonal phase transition arises from a condensation of the $R_{TA}$ mode (antisymmetric octahedral tilts, $R_4^+$), while the tetragonal-to-orthorhombic phase transition is driven by condensation of the $M_{TA_1}$ mode (concerted octahedral tilts, $M_3^+$)~\cite{Benedek:2013jl}. In cesium lead halides the order is reversed~\cite{Fujii:1974di}, likely due to a different coupling mechanism of Cs$^+$ to the $M^+_3$ and $R^+_4$ modes.
The $M^+_3$ and $R^+_4$ distortions may also explain the anomalously large halide atomic displacement
parameters seen in structural analyses of many of these materials~\cite{Worhatch:2008gj,Baikie:2015ju}.

Last, the observation of an instantaneous symmetry broken local structure caused by the combined effects of octahedral tilting and \MA\ and Pb off-centering will
have implications for the electronic band structure, and therefore carrier recombination.
For example, the presence of a local electric polarization can
result in an indirect band gap~\cite{Quarti:2014wp} or spatial separation of the
electron and hole~\cite{Frost:2014dm,Ma:2015he,liu_ferroelectric_2015}, which will reduce carrier
recombination and thus benefit photovoltaic performance.
The off-centering and orientation of the \MA\ ions, coupled to the local
symmetry broken state, can support a quasi-static local polarization that
persists over a wide temperature range if the \MA\ ion displacement and
orientation correlate between crystallographic unit cells. This
polarization may promote the formation of polarons upon photoexcitation,
which can protect carriers from recombination~\cite{Zhu:2015eh}.

\section{Experimental}

\subsection{Sample Preparation}
Single crystals of \MAPI\ were grown via vapor diffusion,\cite{Spingler:2012hx} as reported previously~\cite{Glaser:2015de}. Crystals with original dimensions on the order of 1 mm were polished down to flakes with a thickness of approximately 90 $\mu$m, the X-ray attenuation length of \MAPI\ for an X-ray energy of 23.7 keV. We performed a single-crystal X-ray diffraction experiment to confirm that samples remained single-crystalline after polishing.

\subsection{Data collection}

Measurements were performed on the high-energy resolution inelastic X-ray (HERIX) scattering instrument at Sector 30-ID of the Advanced Photon Source at Argonne National Laboratory with incident beam energy of 23.724 keV ($\lambda=0.5226~\text{\AA}$) and an overall energy resolution of 1.5 meV~\cite{Toellner:mo5010,Said:co5010}. Crystals were mounted on a copper rod using epoxy (Fig.~\ref{fig:xtal}b) and placed inside a beryllium dome. Temperature control was achieved through use of a cryostat. The horizontally polarized incident beam was focused on the sample using a bimorph KB mirror system with a beam size of $15\times35\ \mu \text{m}^2\ (V\times H)$ full width at half maximum (FWHM) at the scattering location. Energy scans, typically in the $\pm8\ \text{meV}$ range with a 0.5 meV step and a collection time of 30 s per point, were taken at fixed momentum transfers $Q = H + q$, where $H$ is the reciprocal lattice vector and $q$ is the phonon wave vector. The scattered beam was analyzed by a system of nine, equally spaced, spherically bent Si(12 12 12) analyzers. The standard momentum transfer resolution of the HERIX instrument is $0.066~\text{\AA}^{-1}$. For the dispersion measurements, we placed a circular slit in front of the analyzer to increase the momentum transfer resolution to $0.020\ \text{\AA}^{-1}$. The basic principles of such instrumentation are discussed elsewhere~\cite{Sinn:2001,Burkel:2001}.

A small elastic component remains at the center at all positions across the zone, coming from static disorder associated with defects in the material. This disorder scattering increases slowly with time in the beam, indicating the presence of beam-damage to the sample (Fig.~S\ref{fig:degradation}). The level of the beam damage is small, containing less intensity than the phonon signals, and was mitigated experimentally by frequently moving the beam to a fresh area of the crystal and realigning the sample.
	
\subsection{Analysis of phonon spectra}

The shape of the incident X-ray energy spectrum was fit using a pseudo-Voigt function. This experimental resolution function $r(\hbar\omega)$ was then convolved with both an elastic and an inelastic scattering component to reproduce the entire spectra as in,
  \begin{equation}
   S(\hbar\omega) = r(\hbar\omega) \ast \left( \delta(\hbar\omega) + F(\hbar\omega) \right),
  \end{equation}
where the elastic component was given by a delta function centered on the zero-point so that the elastic scattering would be given by the resolution function. To model the inelastic scattering component,
a single-phonon scattering model was assumed, and is defined as the response function for a damped harmonic oscillator, given by,

\begin{equation}
	F(\hbar\omega) = N(\hbar\omega)\frac{\Gamma\hbar\omega}{\left(\hbar\omega^2 - \hbar\omega_0^2\right)^2 +\Gamma^2\hbar\omega^2},
\end{equation}
corrected for temperature-dependent occupation of phonon modes and the relationship between energy gain and energy loss processes using a Bose-Einstein distribution adjusted by the detailed balance factor~\cite{dorner1982scattering} $N(\hbar\omega)$:
\begin{equation}
	N(\hbar\omega) = \frac{1}{1 - e^{-\hbar\omega/kT}}.
\end{equation}
Phonon lifetimes were estimated by $\tau = h/\Gamma$. 
For the soft zone-edge modes centered at $\hbar\omega=0$ these are fit with a Lorentzian centered on zero frequency, convoluted with the resolution function, and plotted in Fig.~\ref{fig:dispersion} with the bars of height equal to the half width at half maximum of the Lorentzian.

The respective phonon velocities were extracted from the initial slope of the seven acoustic phonon branches. These phonon velocities were used to extract elastic constants by minimizing the difference between the measured velocities and those predicted by $\mathbf{v}(C_{11},C_{12},C_{44})$. The bulk modulus was calculated from the elastic constants as $K = (C_{11}+2C_{12})/3$.

\subsection{Pair distribution function data collection and analysis}
Total scattering PDF measurements were carried out on beamline 28-ID-2 at the National Synchrotron Light Source II (NSLS-II) at Brookhaven National Laboratory. Data were collected in rapid acquisition mode~\cite{Chupas:wf5000} at an x-ray energy of 67.603~keV ($\lambda=0.18340~\mathrm{\AA}$) and a temperature of 350~K. A large area 2D Perkin-Elmer detector (2048$\times$2048 pixels and 200$\times$200~$\mu$m pixel size) was mounted orthogonal to the beam path with a sample-to-detector distance of 207.5270~mm. Calibration was performed using FIT2D~\cite{Hammersley:1996} on a measurement of nickel. The raw 2D intensity was corrected for experimental effects and azimuthally integrated using FIT2D to obtain the 1D scattering intensity versus the magnitude of the scattering momentum transfer $Q$ ($Q = 4\pi\sin{\theta}/\lambda$ for a scattering angle of 2$\theta$ and x-ray wavelength $\lambda$). xPDFsuite~\cite{Juhas:nb5046, yang2014xpdfsuite} was used for data reduction and Fourier transformation of the total scattering structure function $S(Q)$ to obtain the PDF, $G(r)$, by
\begin{equation}
G(r) = \frac{2}{\pi} \int_{Q_{min}}^{Q_{max}} Q[S(Q)-1]\sin(Qr)dQ,
\end{equation}
where the integration limits, $Q_{min}$--$Q_{max}$, were governed by the experimental setup.

PDFs refinements were carried out using the program PDFgui~\cite{Farrow:2007}, in which PDFs were simulated from model structures using
\begin{equation}
\label{eq:GrModel}
  G(r) = \frac{1}{rN} \sum_{i, j \neq i} \frac{f_{i}^* f_{j}}{\langle f \rangle ^{2}} \delta (r - r_{ij}) - 4\pi r \rho_{0},
\end{equation}
summed over all atoms in the model with periodic boundary conditions on the unit cell. $N$ is the number of atoms, $f_i$ and $f_j$ are the x-ray atomic form factors of atoms $i$ and $j$ respectively, and $\rho_{0}$ is the average atom-pair density. Models were derived from cubic and tetragonal structures of \MAPI\ determined by neutron powder diffraction~\cite{Weller:2015jo}. Unit cell parameters, thermal factors, and symmetry allowed positions were refined to give the best fit to the experimental data. Experimental resolution parameters $Q_{damp}$=0.0434148 and $Q_{broad}$=0.0164506, were determined through refinements of the PDF of the nickel standard.

\subsection{Materials modeling}
First-principles calclations were carried out using the pseudopotential plane-wave density-functional theory (DFT) code, VASP~\cite{Kresse1993}, in conjunction with the Phonopy lattice-dynamics package\cite{Togo2008,Togo2015:jsm}. The calculations are described in detail elsewhere~\cite{Brivio:2015dq}. Projector augmented-wave pseudopotentials~\cite{Blchl1994,Kresse1999} were used, which included the Pb semicore 5d electrons in the valence region. A 700 eV kinetic-energy cutoff was used for the basis set, and a $\Gamma$-centred $k$-point mesh with $6 \times 6 \times 6$ subdivisions was used to sample the electronic Brillouin zone. A tolerance of 10$^{-8}$~eV was applied during the electronic minimisations, and the initial structure was optimised to a force tolerance of 10$^{-3}$~eV/A. These tight convergence criteria were found to be necessary for accurate lattice-dynamics calculations, in particular to eliminate spurious imaginary modes.

Force-constant matrices (FCMs) were calculated from a $2 \times 2 \times 2$  supercell expansion.
Harmonic phonon dispersions were computed along the $\Gamma \rightarrow X$, $\Gamma \rightarrow M$ and $\Gamma \rightarrow R$ segments of the phonon Brillouin zone, as measured in the HERIX experiments, and the nature of the anharmonic modes at $M$ and $R$ were investigated by visualising the phonon-mode eigenvectors (see Supporting Information and \textcolor{blue}{\href{https://figshare.com/s/97a6cbc033b17aa83a18}{Figshare}} for animations). Images of the cubic nonequilibrium and local instantaneous structures (Fig. \ref{fig:pdf}) were generated with VESTA~\cite{Momma2011}. In our model, the \MA\ cation is roughly aligned along the Cartesian \textbf{x} direction, between two faces of the cuboctahedral cavity, which was found in previous work to be the energetically-preferred configuration\cite{Frost:2014ca}. The fixed cation orientation breaks the cubic symmetry, leading to three inequivalent $X$ and $M$ directions, all three of which were analysed in the simulated dispersions.

\begin{acknowledgement}
	Work in the Billinge-group was funded by the US National Science Foundation through grant DMR-1534910. Growth and characterization of single crystals was supported by the Center for Precision Assembly of Superstratic and Superatomic Solids, an NSF MRSEC (Award Number DMR-1420634). This research used resources of the Advanced Photon Source, a U.S. Department of Energy (DOE) Office of Science User Facility operated for the DOE Office of Science by Argonne National Laboratory under Contract No. DE-AC02-06CH11357. X-ray PDF measurements were conducted on beamline 28-ID-2 of the National Synchrotron Light Source II, a U.S. Department of Energy (DOE) Office of Science User Facility operated for the DOE Office of Science by Brookhaven National Laboratory under Contract No. DE-SC0012704.
	The work at Bath has been supported by the EPSRC (Grant Nos. EP/L000202, EP/M009580/1, EP/K016288/1 and EP/K004956/1), and Federico Brivio is thanked for preliminary phonon calculations.
We are grateful to Soham Banerjee for assistance with PDF measurements, and to Bogdan M. Leu, Daniel W. Paley, Ayman Said, John Tranquada, and Omer Yaffe for helpful conversations.	
\end{acknowledgement}

\section{Supporting Information Available:}

Figures showing the crystal of \MAPI\ used for HERIX measurements, HERIX spectra fitting, calculated potential energy surfaces for ion displacement in \MAPI\, comparison of PDF refinements, HERIX energy scans as a function of temperature, and evidence of sample instability in the X-ray beam as well as a table of extracted elastic constants are all available in the Supporting Information.



\bibliography{taviRefs,otherRefs}

\providecommand{\latin}[1]{#1}
\makeatletter
\providecommand{\doi}
  {\begingroup\let\do\@makeother\dospecials
  \catcode`\{=1 \catcode`\}=2\doi@aux}
\providecommand{\doi@aux}[1]{\endgroup\texttt{#1}}
\makeatother
\providecommand*\mcitethebibliography{\thebibliography}
\csname @ifundefined\endcsname{endmcitethebibliography}
  {\let\endmcitethebibliography\endthebibliography}{}
\begin{mcitethebibliography}{57}
\providecommand*\natexlab[1]{#1}
\providecommand*\mciteSetBstSublistMode[1]{}
\providecommand*\mciteSetBstMaxWidthForm[2]{}
\providecommand*\mciteBstWouldAddEndPuncttrue
  {\def\EndOfBibitem{\unskip.}}
\providecommand*\mciteBstWouldAddEndPunctfalse
  {\let\EndOfBibitem\relax}
\providecommand*\mciteSetBstMidEndSepPunct[3]{}
\providecommand*\mciteSetBstSublistLabelBeginEnd[3]{}
\providecommand*\EndOfBibitem{}
\mciteSetBstSublistMode{f}
\mciteSetBstMaxWidthForm{subitem}{(\alph{mcitesubitemcount})}
\mciteSetBstSublistLabelBeginEnd
  {\mcitemaxwidthsubitemform\space}
  {\relax}
  {\relax}

\bibitem[Choi \latin{et~al.}(2014)Choi, Yang, Norman, Billinge, and
  Owen]{Choi:2014ui}
Choi,~J.~J.; Yang,~X.; Norman,~Z.~M.; Billinge,~S. J.~L.; Owen,~J.~S.
  {Structure of Methylammonium Lead Iodide Within Mesoporous Titanium Dioxide:
  Active Material in High-Performance Perovskite Solar Cells}. \emph{Nano
  Lett.} \textbf{2014}, \emph{14}, 127--133\relax
\mciteBstWouldAddEndPuncttrue
\mciteSetBstMidEndSepPunct{\mcitedefaultmidpunct}
{\mcitedefaultendpunct}{\mcitedefaultseppunct}\relax
\EndOfBibitem
\bibitem[Walsh \latin{et~al.}(2015)Walsh, Scanlon, Chen, Gong, and
  Wei]{AronWalsh:2015cq}
Walsh,~A.; Scanlon,~D.~O.; Chen,~S.; Gong,~X.~G.; Wei,~S.-H. {Self-Regulation
  Mechanism for Charged Point Defects in Hybrid Halide Perovskites}.
  \emph{Angew. Chem. Int. Edit.} \textbf{2015}, \emph{54}, 1791--1794\relax
\mciteBstWouldAddEndPuncttrue
\mciteSetBstMidEndSepPunct{\mcitedefaultmidpunct}
{\mcitedefaultendpunct}{\mcitedefaultseppunct}\relax
\EndOfBibitem
\bibitem[Poglitsch and Weber(1987)Poglitsch, and Weber]{Poglitsch:1987ig}
Poglitsch,~A.; Weber,~D. {Dynamic disorder in
  methylammoniumtrihalogenoplumbates (II) observed by millimeter-wave
  spectroscopy}. \emph{J. Chem. Phys.} \textbf{1987}, \emph{87},
  6373--6378\relax
\mciteBstWouldAddEndPuncttrue
\mciteSetBstMidEndSepPunct{\mcitedefaultmidpunct}
{\mcitedefaultendpunct}{\mcitedefaultseppunct}\relax
\EndOfBibitem
\bibitem[Yaffe \latin{et~al.}(2016)Yaffe, Guo, Hull, Stoumpos, Tan, Egger,
  Zheng, Szpak, Semonin, Beecher, Heinz, Kronik, Rappe, Kanatzidis, Owen,
  Pimenta, and Brus]{Yaffe:2016wt}
Yaffe,~O.; Guo,~Y.; Hull,~T.; Stoumpos,~C.~C.; Tan,~L.~Z.; Egger,~D.~A.;
  Zheng,~F.; Szpak,~.~G.; Semonin,~O.~E.; Beecher,~A.~N. \latin{et~al.}  The
  nature of dynamic disorder in lead halide perovskite crystals.
  \emph{arXiv:1604.08107 [cond-mat]} \textbf{2016}, arXiv: 1604.08107\relax
\mciteBstWouldAddEndPuncttrue
\mciteSetBstMidEndSepPunct{\mcitedefaultmidpunct}
{\mcitedefaultendpunct}{\mcitedefaultseppunct}\relax
\EndOfBibitem
\bibitem[Stranks and Snaith(2015)Stranks, and Snaith]{Stranks:2015ef}
Stranks,~S.~D.; Snaith,~H.~J. {Metal-halide perovskites for photovoltaic and
  light-emitting devices}. \emph{Nat. Nanotechnol.} \textbf{2015}, \emph{10},
  391--402\relax
\mciteBstWouldAddEndPuncttrue
\mciteSetBstMidEndSepPunct{\mcitedefaultmidpunct}
{\mcitedefaultendpunct}{\mcitedefaultseppunct}\relax
\EndOfBibitem
\bibitem[Green \latin{et~al.}(2016)Green, Emery, Hishikawa, Warta, and
  Dunlop]{Green:2016fj}
Green,~M.~A.; Emery,~K.; Hishikawa,~Y.; Warta,~W.; Dunlop,~E.~D. {Solar cell
  efficiency tables (version 47)}. \emph{Prog. Photovoltaics} \textbf{2016},
  \emph{24}, 3--11\relax
\mciteBstWouldAddEndPuncttrue
\mciteSetBstMidEndSepPunct{\mcitedefaultmidpunct}
{\mcitedefaultendpunct}{\mcitedefaultseppunct}\relax
\EndOfBibitem
\bibitem[Glazer(1975)]{Glazer:1975er}
Glazer,~A.~M. {Simple ways of determining perovskite structures}. \emph{Acta
  Crystallogr., Sect. A} \textbf{1975}, \emph{31}, 756--762\relax
\mciteBstWouldAddEndPuncttrue
\mciteSetBstMidEndSepPunct{\mcitedefaultmidpunct}
{\mcitedefaultendpunct}{\mcitedefaultseppunct}\relax
\EndOfBibitem
\bibitem[Benedek and Fennie(2013)Benedek, and Fennie]{Benedek:2013jl}
Benedek,~N.~A.; Fennie,~C.~J. {Why Are There So Few Perovskite Ferroelectrics?}
  \emph{J. Phys. Chem. C} \textbf{2013}, \emph{117}, 13339--13349\relax
\mciteBstWouldAddEndPuncttrue
\mciteSetBstMidEndSepPunct{\mcitedefaultmidpunct}
{\mcitedefaultendpunct}{\mcitedefaultseppunct}\relax
\EndOfBibitem
\bibitem[Frost \latin{et~al.}(2014)Frost, Butler, Brivio, Hendon, van
  Schilfgaarde, and Walsh]{Frost:2014dm}
Frost,~J.~M.; Butler,~K.~T.; Brivio,~F.; Hendon,~C.~H.; van Schilfgaarde,~M.;
  Walsh,~A. {Atomistic Origins of High-Performance in Hybrid Halide Perovskite
  Solar Cells}. \emph{Nano Lett.} \textbf{2014}, \emph{14}, 2584--2590\relax
\mciteBstWouldAddEndPuncttrue
\mciteSetBstMidEndSepPunct{\mcitedefaultmidpunct}
{\mcitedefaultendpunct}{\mcitedefaultseppunct}\relax
\EndOfBibitem
\bibitem[Ma and Wang(2015)Ma, and Wang]{Ma:2015he}
Ma,~J.; Wang,~L.-W. {Nanoscale Charge Localization Induced by Random
  Orientations of Organic Molecules in Hybrid Perovskite CH 3NH 3PbI 3}.
  \emph{Nano Lett.} \textbf{2015}, \emph{15}, 248--253\relax
\mciteBstWouldAddEndPuncttrue
\mciteSetBstMidEndSepPunct{\mcitedefaultmidpunct}
{\mcitedefaultendpunct}{\mcitedefaultseppunct}\relax
\EndOfBibitem
\bibitem[Zhu and Podzorov(2015)Zhu, and Podzorov]{Zhu:2015eh}
Zhu,~X.~Y.; Podzorov,~V. {Charge Carriers in Hybrid Organic--Inorganic Lead
  Halide Perovskites Might Be Protected as Large Polarons}. \emph{J. Phys.
  Chem. Lett.} \textbf{2015}, \emph{6}, 4758--4761\relax
\mciteBstWouldAddEndPuncttrue
\mciteSetBstMidEndSepPunct{\mcitedefaultmidpunct}
{\mcitedefaultendpunct}{\mcitedefaultseppunct}\relax
\EndOfBibitem
\bibitem[Liu \latin{et~al.}(2015)Liu, Zheng, Koocher, Takenaka, Wang, and
  Rappe]{liu_ferroelectric_2015}
Liu,~S.; Zheng,~F.; Koocher,~N.~Z.; Takenaka,~H.; Wang,~F.; Rappe,~A.~M.
  Ferroelectric {Domain} {Wall} {Induced} {Band} {Gap} {Reduction} and {Charge}
  {Separation} in {Organometal} {Halide} {Perovskites}. \emph{J. Phys. Chem.
  Lett.} \textbf{2015}, \emph{6}, 693--699\relax
\mciteBstWouldAddEndPuncttrue
\mciteSetBstMidEndSepPunct{\mcitedefaultmidpunct}
{\mcitedefaultendpunct}{\mcitedefaultseppunct}\relax
\EndOfBibitem
\bibitem[Stoumpos \latin{et~al.}(2013)Stoumpos, Malliakas, and
  Kanatzidis]{Stoumpos:2013wq}
Stoumpos,~C.~C.; Malliakas,~C.~D.; Kanatzidis,~M.~G. {Semiconducting tin and
  lead iodide perovskites with organic cations: phase transitions, high
  mobilities, and near-infrared photoluminescent properties}. \emph{Inorg.
  Chem.} \textbf{2013}, \emph{52}, 9019--9038\relax
\mciteBstWouldAddEndPuncttrue
\mciteSetBstMidEndSepPunct{\mcitedefaultmidpunct}
{\mcitedefaultendpunct}{\mcitedefaultseppunct}\relax
\EndOfBibitem
\bibitem[Baikie \latin{et~al.}(2015)Baikie, Barrow, Fang, Keenan, Slater,
  Piltz, Gutmann, Mhaisalkar, and White]{Baikie:2015ju}
Baikie,~T.; Barrow,~N.~S.; Fang,~Y.; Keenan,~P.~J.; Slater,~P.~R.;
  Piltz,~R.~O.; Gutmann,~M.; Mhaisalkar,~S.~G.; White,~T.~J. {A combined single
  crystal neutron/X-ray diffraction and solid-state nuclear magnetic resonance
  study of the hybrid perovskites CH 3NH 3PbX 3(X = I, Br and Cl)}. \emph{J.
  Mater. Chem. A} \textbf{2015}, \emph{3}, 9298--9307\relax
\mciteBstWouldAddEndPuncttrue
\mciteSetBstMidEndSepPunct{\mcitedefaultmidpunct}
{\mcitedefaultendpunct}{\mcitedefaultseppunct}\relax
\EndOfBibitem
\bibitem[Stroppa \latin{et~al.}(2015)Stroppa, Quarti, De~Angelis, and
  Picozzi]{stroppa_ferroelectric_2015}
Stroppa,~A.; Quarti,~C.; De~Angelis,~F.; Picozzi,~S. Ferroelectric
  {Polarization} of {CH}3NH3PbI3: {A} {Detailed} {Study} {Based} on {Density}
  {Functional} {Theory} and {Symmetry} {Mode} {Analysis}. \emph{J. Phys. Chem.
  Lett.} \textbf{2015}, \emph{6}, 2223--2231\relax
\mciteBstWouldAddEndPuncttrue
\mciteSetBstMidEndSepPunct{\mcitedefaultmidpunct}
{\mcitedefaultendpunct}{\mcitedefaultseppunct}\relax
\EndOfBibitem
\bibitem[Beilsten-Edmands \latin{et~al.}(2015)Beilsten-Edmands, Eperon,
  Johnson, Snaith, and Radaelli]{beilsten-edmands_non-ferroelectric_2015}
Beilsten-Edmands,~J.; Eperon,~G.~E.; Johnson,~R.~D.; Snaith,~H.~J.;
  Radaelli,~P.~G. Non-ferroelectric nature of the conductance hysteresis in
  {CH}3NH3PbI3 perovskite-based photovoltaic devices. \emph{Appl. Phys. Lett.}
  \textbf{2015}, \emph{106}, 173502\relax
\mciteBstWouldAddEndPuncttrue
\mciteSetBstMidEndSepPunct{\mcitedefaultmidpunct}
{\mcitedefaultendpunct}{\mcitedefaultseppunct}\relax
\EndOfBibitem
\bibitem[Wasylishen \latin{et~al.}(1985)Wasylishen, Knop, and
  Macdonald]{wasylishen_cation_1985}
Wasylishen,~R.~E.; Knop,~O.; Macdonald,~J.~B. Cation rotation in methylammonium
  lead halides. \emph{Solid State Comm.} \textbf{1985}, \emph{56},
  581--582\relax
\mciteBstWouldAddEndPuncttrue
\mciteSetBstMidEndSepPunct{\mcitedefaultmidpunct}
{\mcitedefaultendpunct}{\mcitedefaultseppunct}\relax
\EndOfBibitem
\bibitem[Leguy \latin{et~al.}(2015)Leguy, Frost, McMahon, Sakai, Kochelmann,
  Law, Li, Foglia, Walsh, O'Regan, Nelson, Cabral, and Barnes]{Leguy:2015hq}
Leguy,~A. M.~A.; Frost,~J.~M.; McMahon,~A.~P.; Sakai,~V.~G.; Kochelmann,~W.;
  Law,~C.; Li,~X.; Foglia,~F.; Walsh,~A.; O'Regan,~B.~C. \latin{et~al.}  {The
  dynamics of methylammonium ions in hybrid organic--inorganic perovskite solar
  cells}. \emph{Nat. Commun.} \textbf{2015}, \emph{6}, 7124\relax
\mciteBstWouldAddEndPuncttrue
\mciteSetBstMidEndSepPunct{\mcitedefaultmidpunct}
{\mcitedefaultendpunct}{\mcitedefaultseppunct}\relax
\EndOfBibitem
\bibitem[Chen \latin{et~al.}(2015)Chen, Foley, Ipek, Tyagi, Copley, Brown,
  Choi, and Lee]{Chen:2015jp}
Chen,~T.; Foley,~B.~J.; Ipek,~B.; Tyagi,~M.; Copley,~J. R.~D.; Brown,~C.~M.;
  Choi,~J.~J.; Lee,~S.-H. {Rotational dynamics of organic cations in the CH 3
  NH 3 PbI 3 perovskite}. \emph{Phys. Chem. Chem. Phys.} \textbf{2015},
  \emph{17}, 31278--31286\relax
\mciteBstWouldAddEndPuncttrue
\mciteSetBstMidEndSepPunct{\mcitedefaultmidpunct}
{\mcitedefaultendpunct}{\mcitedefaultseppunct}\relax
\EndOfBibitem
\bibitem[Quarti \latin{et~al.}(2014)Quarti, Mosconi, and
  De~Angelis]{Quarti:2014wp}
Quarti,~C.; Mosconi,~E.; De~Angelis,~F. Interplay of Orientational Order and
  Electronic Structure in Methylammonium Lead Iodide: Implications for Solar
  Cell Operation. \emph{Chem. Mater.} \textbf{2014}, \emph{26},
  6557--6569\relax
\mciteBstWouldAddEndPuncttrue
\mciteSetBstMidEndSepPunct{\mcitedefaultmidpunct}
{\mcitedefaultendpunct}{\mcitedefaultseppunct}\relax
\EndOfBibitem
\bibitem[Brivio \latin{et~al.}(2015)Brivio, Frost, Skelton, Jackson, Weber,
  Weller, Go{\~n}i, Leguy, Barnes, and Walsh]{Brivio:2015dq}
Brivio,~F.; Frost,~J.~M.; Skelton,~J.~M.; Jackson,~A.~J.; Weber,~O.~J.;
  Weller,~M.~T.; Go{\~n}i,~A.~R.; Leguy,~A. M.~A.; Barnes,~P. R.~F.; Walsh,~A.
  {Lattice dynamics and vibrational spectra of the orthorhombic, tetragonal,
  and cubic phases of methylammonium lead iodide}. \emph{Phys. Rev. B}
  \textbf{2015}, \emph{92}, 144308\relax
\mciteBstWouldAddEndPuncttrue
\mciteSetBstMidEndSepPunct{\mcitedefaultmidpunct}
{\mcitedefaultendpunct}{\mcitedefaultseppunct}\relax
\EndOfBibitem
\bibitem[Quarti \latin{et~al.}(2016)Quarti, Mosconi, Ball, D'Innocenzo, Tao,
  Pathak, Snaith, Petrozza, and De~Angelis]{quarti_structural_2016}
Quarti,~C.; Mosconi,~E.; Ball,~J.~M.; D'Innocenzo,~V.; Tao,~C.; Pathak,~S.;
  Snaith,~H.~J.; Petrozza,~A.; De~Angelis,~F. Structural and optical properties
  of methylammonium lead iodide across the tetragonal to cubic phase
  transition: implications for perovskite solar cells. \emph{Energy Environ.
  Sci.} \textbf{2016}, \emph{9}, 155--163\relax
\mciteBstWouldAddEndPuncttrue
\mciteSetBstMidEndSepPunct{\mcitedefaultmidpunct}
{\mcitedefaultendpunct}{\mcitedefaultseppunct}\relax
\EndOfBibitem
\bibitem[Fujii \latin{et~al.}(1974)Fujii, Hoshino, Yamada, and
  Shirane]{Fujii:1974di}
Fujii,~Y.; Hoshino,~S.; Yamada,~Y.; Shirane,~G. {Neutron-scattering study on
  phase transitions of CsPb Cl3}. \emph{Phys. Rev. B} \textbf{1974}, \emph{9},
  4549--4559\relax
\mciteBstWouldAddEndPuncttrue
\mciteSetBstMidEndSepPunct{\mcitedefaultmidpunct}
{\mcitedefaultendpunct}{\mcitedefaultseppunct}\relax
\EndOfBibitem
\bibitem[Swainson \latin{et~al.}(2003)Swainson, Hammond, Soulli{\`e}re, Knop,
  and Massa]{Swainson:2003hy}
Swainson,~I.~P.; Hammond,~R.~P.; Soulli{\`e}re,~C.; Knop,~O.; Massa,~W. {Phase
  transitions in the perovskite methylammonium lead bromide, CH3ND3PbBr3}.
  \emph{J. Solid State Chem.} \textbf{2003}, \emph{176}, 97--104\relax
\mciteBstWouldAddEndPuncttrue
\mciteSetBstMidEndSepPunct{\mcitedefaultmidpunct}
{\mcitedefaultendpunct}{\mcitedefaultseppunct}\relax
\EndOfBibitem
\bibitem[Chi \latin{et~al.}(2005)Chi, Swainson, Cranswick, Her, Stephens, and
  Knop]{Chi:2005if}
Chi,~L.; Swainson,~I.; Cranswick,~L.; Her,~J.-H.; Stephens,~P.; Knop,~O. {The
  ordered phase of methylammonium lead chloride CH3ND3PbCl3}. \emph{J. Solid
  State Chem.} \textbf{2005}, \emph{178}, 1376--1385\relax
\mciteBstWouldAddEndPuncttrue
\mciteSetBstMidEndSepPunct{\mcitedefaultmidpunct}
{\mcitedefaultendpunct}{\mcitedefaultseppunct}\relax
\EndOfBibitem
\bibitem[Swainson \latin{et~al.}(2015)Swainson, Stock, Parker, Van~Eijck,
  Russina, and Taylor]{Swainson:2015ez}
Swainson,~I.~P.; Stock,~C.; Parker,~S.~F.; Van~Eijck,~L.; Russina,~M.;
  Taylor,~J.~W. {From soft harmonic phonons to fast relaxational dynamics in
  CH3NH3PbBr3}. \emph{Phys. Rev. B} \textbf{2015}, \emph{92}, 100303\relax
\mciteBstWouldAddEndPuncttrue
\mciteSetBstMidEndSepPunct{\mcitedefaultmidpunct}
{\mcitedefaultendpunct}{\mcitedefaultseppunct}\relax
\EndOfBibitem
\bibitem[Gindl and Gupta(2002)Gindl, and Gupta]{Gindl:2002il}
Gindl,~W.; Gupta,~H.~S. {Cell-wall hardness and Young's modulus of
  melamine-modified spruce wood by nano-indentation}. \emph{Composites, Part A}
  \textbf{2002}, \emph{33}, 1141--1145\relax
\mciteBstWouldAddEndPuncttrue
\mciteSetBstMidEndSepPunct{\mcitedefaultmidpunct}
{\mcitedefaultendpunct}{\mcitedefaultseppunct}\relax
\EndOfBibitem
\bibitem[Pisoni \latin{et~al.}(2014)Pisoni, Ja{\'c}imovi{\'c}, Bari{\v
  s}i{\'c}, Spina, Gaal, Forr{\'o}, and Horv{\'a}th]{Pisoni:2014jy}
Pisoni,~A.; Ja{\'c}imovi{\'c},~J.; Bari{\v s}i{\'c},~O.~S.; Spina,~M.;
  Gaal,~R.; Forr{\'o},~L.; Horv{\'a}th,~E. {Ultra-Low Thermal Conductivity in
  Organic--Inorganic Hybrid Perovskite CH 3NH 3PbI 3}. \emph{J. Phys. Chem.
  Lett.} \textbf{2014}, \emph{5}, 2488--2492\relax
\mciteBstWouldAddEndPuncttrue
\mciteSetBstMidEndSepPunct{\mcitedefaultmidpunct}
{\mcitedefaultendpunct}{\mcitedefaultseppunct}\relax
\EndOfBibitem
\bibitem[Delaire \latin{et~al.}(2011)Delaire, Ma, Marty, May, McGuire, Du,
  Singh, Podlesnyak, Ehlers, Lumsden, and Sales]{Delaire:2011gr}
Delaire,~O.; Ma,~J.; Marty,~K.; May,~A.~F.; McGuire,~M.~A.; Du,~M.-H.;
  Singh,~D.~J.; Podlesnyak,~A.; Ehlers,~G.; Lumsden,~M.~D. \latin{et~al.}
  {Giant anharmonic phonon scattering in PbTe}. \emph{Nat. Mater.}
  \textbf{2011}, \emph{10}, 614--619\relax
\mciteBstWouldAddEndPuncttrue
\mciteSetBstMidEndSepPunct{\mcitedefaultmidpunct}
{\mcitedefaultendpunct}{\mcitedefaultseppunct}\relax
\EndOfBibitem
\bibitem[Li \latin{et~al.}(2015)Li, Hong, May, Bansal, Chi, Hong, Ehlers, and
  Delaire]{Li:2015gf}
Li,~C.~W.; Hong,~J.; May,~A.~F.; Bansal,~D.; Chi,~S.; Hong,~T.; Ehlers,~G.;
  Delaire,~O. {Orbitally driven giant phonon anharmonicity in SnSe}. \emph{Nat.
  Phys.} \textbf{2015}, \emph{11}, 1063--1069\relax
\mciteBstWouldAddEndPuncttrue
\mciteSetBstMidEndSepPunct{\mcitedefaultmidpunct}
{\mcitedefaultendpunct}{\mcitedefaultseppunct}\relax
\EndOfBibitem
\bibitem[Karakus \latin{et~al.}(2015)Karakus, Jensen, D'Angelo, Turchinovich,
  Bonn, and C{\'a}novas]{Karakus:2015de}
Karakus,~M.; Jensen,~S.~A.; D'Angelo,~F.; Turchinovich,~D.; Bonn,~M.;
  C{\'a}novas,~E. {Phonon--Electron Scattering Limits Free Charge Mobility in
  Methylammonium Lead Iodide Perovskites}. \emph{J. Phys. Chem. Lett.}
  \textbf{2015}, \emph{6}, 4991--4996\relax
\mciteBstWouldAddEndPuncttrue
\mciteSetBstMidEndSepPunct{\mcitedefaultmidpunct}
{\mcitedefaultendpunct}{\mcitedefaultseppunct}\relax
\EndOfBibitem
\bibitem[Frost and Walsh(2016)Frost, and Walsh]{Frost:2016kl}
Frost,~J.~M.; Walsh,~A. {What Is Moving in Hybrid Halide Perovskite Solar
  Cells?} \emph{Acc. Chem. Res.} \textbf{2016}, \emph{49}, 528--535\relax
\mciteBstWouldAddEndPuncttrue
\mciteSetBstMidEndSepPunct{\mcitedefaultmidpunct}
{\mcitedefaultendpunct}{\mcitedefaultseppunct}\relax
\EndOfBibitem
\bibitem[Weller \latin{et~al.}(2015)Weller, Weber, Henry, Di~Pumpo, and
  Hansen]{Weller:2015jo}
Weller,~M.~T.; Weber,~O.~J.; Henry,~P.~F.; Di~Pumpo,~A.~M.; Hansen,~T.~C.
  {Complete structure and cation orientation in the perovskite photovoltaic
  methylammonium lead iodide between 100 and 352 K }. \emph{Chem. Commun.}
  \textbf{2015}, \emph{51}, 4180--4183\relax
\mciteBstWouldAddEndPuncttrue
\mciteSetBstMidEndSepPunct{\mcitedefaultmidpunct}
{\mcitedefaultendpunct}{\mcitedefaultseppunct}\relax
\EndOfBibitem
\bibitem[Ren \latin{et~al.}(2016)Ren, Oswald, Wang, McCandless, and
  Chan]{Ren:2016cw}
Ren,~Y.; Oswald,~I. W.~H.; Wang,~X.; McCandless,~G.~T.; Chan,~J.~Y.
  {Orientation of Organic Cations in Hybrid Inorganic--Organic Perovskite
  CH$_3$NH$_3$PbI$_3$ from Subatomic Resolution Single Crystal Neutron
  Diffraction Structural Studies}. \emph{Cryst. Growth Des.} \textbf{2016},
  \emph{16}, 2945--2951\relax
\mciteBstWouldAddEndPuncttrue
\mciteSetBstMidEndSepPunct{\mcitedefaultmidpunct}
{\mcitedefaultendpunct}{\mcitedefaultseppunct}\relax
\EndOfBibitem
\bibitem[Brivio \latin{et~al.}(2014)Brivio, Butler, Walsh, and van
  Schilfgaarde]{Brivio2014.PRB}
Brivio,~F.; Butler,~K.~T.; Walsh,~A.; van Schilfgaarde,~M. Relativistic
  quasiparticle self-consistent electronic structure of hybrid halide
  perovskite photovoltaic absorbers. \emph{Phys. Rev. B} \textbf{2014},
  \emph{89}, 155204\relax
\mciteBstWouldAddEndPuncttrue
\mciteSetBstMidEndSepPunct{\mcitedefaultmidpunct}
{\mcitedefaultendpunct}{\mcitedefaultseppunct}\relax
\EndOfBibitem
\bibitem[Kocsis(1975)]{Kocsis1975}
Kocsis,~S. Lattice scattering mobility of electrons in {GaP}. \emph{Phys.
  Status Solidi A} \textbf{1975}, \emph{28}, 133--138\relax
\mciteBstWouldAddEndPuncttrue
\mciteSetBstMidEndSepPunct{\mcitedefaultmidpunct}
{\mcitedefaultendpunct}{\mcitedefaultseppunct}\relax
\EndOfBibitem
\bibitem[Worhatch \latin{et~al.}(2008)Worhatch, Kim, Swainson, Yonkeu, and
  Billinge]{Worhatch:2008gj}
Worhatch,~R.~J.; Kim,~H.; Swainson,~I.~P.; Yonkeu,~A.~L.; Billinge,~S. J.~L.
  {Study of Local Structure in Selected Organic--Inorganic Perovskites in the
  Pm3m Phase}. \emph{Chem. Mater.} \textbf{2008}, \emph{20}, 1272--1277\relax
\mciteBstWouldAddEndPuncttrue
\mciteSetBstMidEndSepPunct{\mcitedefaultmidpunct}
{\mcitedefaultendpunct}{\mcitedefaultseppunct}\relax
\EndOfBibitem
\bibitem[Spingler \latin{et~al.}(2012)Spingler, Schnidrig, Todorova, and
  Wild]{Spingler:2012hx}
Spingler,~B.; Schnidrig,~S.; Todorova,~T.; Wild,~F. {Some thoughts about the
  single crystal growth of small molecules }. \emph{Cryst. Eng. Comm.}
  \textbf{2012}, \emph{14}, 751--757\relax
\mciteBstWouldAddEndPuncttrue
\mciteSetBstMidEndSepPunct{\mcitedefaultmidpunct}
{\mcitedefaultendpunct}{\mcitedefaultseppunct}\relax
\EndOfBibitem
\bibitem[Glaser \latin{et~al.}(2015)Glaser, M{\"u}ller, Sendner, Krekeler,
  Semonin, Hull, Yaffe, Owen, Kowalsky, Pucci, and Lovrin{\v
  c}i{\'c}]{Glaser:2015de}
Glaser,~T.; M{\"u}ller,~C.; Sendner,~M.; Krekeler,~C.; Semonin,~O.~E.;
  Hull,~T.~D.; Yaffe,~O.; Owen,~J.~S.; Kowalsky,~W.; Pucci,~A. \latin{et~al.}
  {Infrared Spectroscopic Study of Vibrational Modes in Methylammonium Lead
  Halide Perovskites}. \emph{J. Phys. Chem. Lett.} \textbf{2015}, \emph{6},
  2913--2918\relax
\mciteBstWouldAddEndPuncttrue
\mciteSetBstMidEndSepPunct{\mcitedefaultmidpunct}
{\mcitedefaultendpunct}{\mcitedefaultseppunct}\relax
\EndOfBibitem
\bibitem[Toellner \latin{et~al.}(2011)Toellner, Alatas, and
  Said]{Toellner:mo5010}
Toellner,~T.~S.; Alatas,~A.; Said,~A.~H. {Six-reflection meV-monochromator for
  synchrotron radiation}. \emph{J. Synchrotron Radiat.} \textbf{2011},
  \emph{18}, 605--611\relax
\mciteBstWouldAddEndPuncttrue
\mciteSetBstMidEndSepPunct{\mcitedefaultmidpunct}
{\mcitedefaultendpunct}{\mcitedefaultseppunct}\relax
\EndOfBibitem
\bibitem[Said \latin{et~al.}(2011)Said, Sinn, and Divan]{Said:co5010}
Said,~A.~H.; Sinn,~H.; Divan,~R. {New developments in fabrication of
  high-energy-resolution analyzers for inelastic X-ray spectroscopy}. \emph{J.
  Synchrotron Radiat.} \textbf{2011}, \emph{18}, 492--496\relax
\mciteBstWouldAddEndPuncttrue
\mciteSetBstMidEndSepPunct{\mcitedefaultmidpunct}
{\mcitedefaultendpunct}{\mcitedefaultseppunct}\relax
\EndOfBibitem
\bibitem[Sinn(2001)]{Sinn:2001}
Sinn,~H. Spectroscopy with meV energy resolution. \emph{J. Phys.: Condens.
  Matter} \textbf{2001}, \emph{13}, 7525--7537\relax
\mciteBstWouldAddEndPuncttrue
\mciteSetBstMidEndSepPunct{\mcitedefaultmidpunct}
{\mcitedefaultendpunct}{\mcitedefaultseppunct}\relax
\EndOfBibitem
\bibitem[Burkel(2000)]{Burkel:2001}
Burkel,~E. Phonon spectroscopy by inelastic x-ray scattering. \emph{Rep. Prog.
  Phys.} \textbf{2000}, \emph{63}, 171--232\relax
\mciteBstWouldAddEndPuncttrue
\mciteSetBstMidEndSepPunct{\mcitedefaultmidpunct}
{\mcitedefaultendpunct}{\mcitedefaultseppunct}\relax
\EndOfBibitem
\bibitem[Dorner(1982)]{dorner1982scattering}
Dorner,~B. The scattering function and symmetry operations in the crystal.
  \emph{Coherent inelastic neutron scaterring in lattice dynamics}
  \textbf{1982}, 16--24\relax
\mciteBstWouldAddEndPuncttrue
\mciteSetBstMidEndSepPunct{\mcitedefaultmidpunct}
{\mcitedefaultendpunct}{\mcitedefaultseppunct}\relax
\EndOfBibitem
\bibitem[Chupas \latin{et~al.}(2003)Chupas, Qiu, Hanson, Lee, Grey, and
  Billinge]{Chupas:wf5000}
Chupas,~P.~J.; Qiu,~X.; Hanson,~J.~C.; Lee,~P.~L.; Grey,~C.~P.; Billinge,~S.
  J.~L. {Rapid-acquisition pair distribution function (RA-PDF) analysis}.
  \emph{J. Appl. Crystallogr.} \textbf{2003}, \emph{36}, 1342--1347\relax
\mciteBstWouldAddEndPuncttrue
\mciteSetBstMidEndSepPunct{\mcitedefaultmidpunct}
{\mcitedefaultendpunct}{\mcitedefaultseppunct}\relax
\EndOfBibitem
\bibitem[Hammersley \latin{et~al.}(1996)Hammersley, Svensson, Hanfland, Fitch,
  and Hausermann]{Hammersley:1996}
Hammersley,~A.~P.; Svensson,~S.~O.; Hanfland,~M.; Fitch,~A.~N.; Hausermann,~D.
  Two-dimensional detector software: From real detector to idealised image or
  two-theta scan. \emph{High Pressure Res.} \textbf{1996}, \emph{14},
  235--248\relax
\mciteBstWouldAddEndPuncttrue
\mciteSetBstMidEndSepPunct{\mcitedefaultmidpunct}
{\mcitedefaultendpunct}{\mcitedefaultseppunct}\relax
\EndOfBibitem
\bibitem[Juh{\'{a}}s \latin{et~al.}(2013)Juh{\'{a}}s, Davis, Farrow, and
  Billinge]{Juhas:nb5046}
Juh{\'{a}}s,~P.; Davis,~T.; Farrow,~C.~L.; Billinge,~S.~J.~L. {{\it PDFgetX3}:
  a rapid and highly automatable program for processing powder diffraction data
  into total scattering pair distribution functions}. \emph{J. Appl.
  Crystallogr.} \textbf{2013}, \emph{46}, 560--566\relax
\mciteBstWouldAddEndPuncttrue
\mciteSetBstMidEndSepPunct{\mcitedefaultmidpunct}
{\mcitedefaultendpunct}{\mcitedefaultseppunct}\relax
\EndOfBibitem
\bibitem[Yang \latin{et~al.}(2014)Yang, Juh{\'{a}}s, Farrow, and
  Billinge]{yang2014xpdfsuite}
Yang,~X.; Juh{\'{a}}s,~P.; Farrow,~C.~L.; Billinge,~S. J.~L. {xPDFsuite}: an
  end-to-end software solution for high throughput pair distribution function
  transformation, visualization and analysis. \emph{arXiv:1402.3163 [cond-mat]}
  \textbf{2014}, arXiv: 1402.3163\relax
\mciteBstWouldAddEndPuncttrue
\mciteSetBstMidEndSepPunct{\mcitedefaultmidpunct}
{\mcitedefaultendpunct}{\mcitedefaultseppunct}\relax
\EndOfBibitem
\bibitem[Farrow \latin{et~al.}(2007)Farrow, Juhas, Liu, Bryndin, Bo{\v{z}}in,
  Bloch, Proffen, and Billinge]{Farrow:2007}
Farrow,~C.~L.; Juhas,~P.; Liu,~J.~W.; Bryndin,~D.; Bo{\v{z}}in,~E.~S.;
  Bloch,~J.; Proffen,~T.; Billinge,~S. J.~L. PDFfit2 and PDFgui: computer
  programs for studying nanostructure in crystals. \emph{J. Phys.: Condens.
  Matter} \textbf{2007}, \emph{19}, 335219\relax
\mciteBstWouldAddEndPuncttrue
\mciteSetBstMidEndSepPunct{\mcitedefaultmidpunct}
{\mcitedefaultendpunct}{\mcitedefaultseppunct}\relax
\EndOfBibitem
\bibitem[Kresse and Hafner(1993)Kresse, and Hafner]{Kresse1993}
Kresse,~G.; Hafner,~J. Ab initio molecular dynamics for liquid metals.
  \emph{Phys. Rev. B} \textbf{1993}, \emph{47}, 558--561\relax
\mciteBstWouldAddEndPuncttrue
\mciteSetBstMidEndSepPunct{\mcitedefaultmidpunct}
{\mcitedefaultendpunct}{\mcitedefaultseppunct}\relax
\EndOfBibitem
\bibitem[Togo \latin{et~al.}(2008)Togo, Oba, and Tanaka]{Togo2008}
Togo,~A.; Oba,~F.; Tanaka,~I. First-principles calculations of the ferroelastic
  transition between rutile-type and CaCl$_2$-type SiO$_2$ at high pressures.
  \emph{Phys. Rev. B} \textbf{2008}, \emph{78}, 134106\relax
\mciteBstWouldAddEndPuncttrue
\mciteSetBstMidEndSepPunct{\mcitedefaultmidpunct}
{\mcitedefaultendpunct}{\mcitedefaultseppunct}\relax
\EndOfBibitem
\bibitem[Togo and Tanaka(2015)Togo, and Tanaka]{Togo2015:jsm}
Togo,~A.; Tanaka,~I. First principles phonon calculations in materials science.
  \emph{Scr. Mater.} \textbf{2015}, \emph{108}, 1--5\relax
\mciteBstWouldAddEndPuncttrue
\mciteSetBstMidEndSepPunct{\mcitedefaultmidpunct}
{\mcitedefaultendpunct}{\mcitedefaultseppunct}\relax
\EndOfBibitem
\bibitem[Bl\"{o}chl(1994)]{Blchl1994}
Bl\"{o}chl,~P.~E. Projector augmented-wave method. \emph{Phys. Rev. B}
  \textbf{1994}, \emph{50}, 17953--17979\relax
\mciteBstWouldAddEndPuncttrue
\mciteSetBstMidEndSepPunct{\mcitedefaultmidpunct}
{\mcitedefaultendpunct}{\mcitedefaultseppunct}\relax
\EndOfBibitem
\bibitem[Kresse and Joubert(1999)Kresse, and Joubert]{Kresse1999}
Kresse,~G.; Joubert,~D. From ultrasoft pseudopotentials to the projector
  augmented-wave method. \emph{Phys. Rev. B} \textbf{1999}, \emph{59},
  1758--1775\relax
\mciteBstWouldAddEndPuncttrue
\mciteSetBstMidEndSepPunct{\mcitedefaultmidpunct}
{\mcitedefaultendpunct}{\mcitedefaultseppunct}\relax
\EndOfBibitem
\bibitem[Momma and Izumi(2011)Momma, and Izumi]{Momma2011}
Momma,~K.; Izumi,~F. {VESTA} 3 for three-dimensional visualization of crystal,
  volumetric and morphology data. \emph{J. Appl. Cryst.} \textbf{2011},
  \emph{44}, 1272--1276\relax
\mciteBstWouldAddEndPuncttrue
\mciteSetBstMidEndSepPunct{\mcitedefaultmidpunct}
{\mcitedefaultendpunct}{\mcitedefaultseppunct}\relax
\EndOfBibitem
\bibitem[Frost \latin{et~al.}(2014)Frost, Butler, and Walsh]{Frost:2014ca}
Frost,~J.~M.; Butler,~K.~T.; Walsh,~A. {Molecular ferroelectric contributions
  to anomalous hysteresis in hybrid perovskite solar cells}. \emph{APL Mater.}
  \textbf{2014}, \emph{2}, 081506\relax
\mciteBstWouldAddEndPuncttrue
\mciteSetBstMidEndSepPunct{\mcitedefaultmidpunct}
{\mcitedefaultendpunct}{\mcitedefaultseppunct}\relax
\EndOfBibitem
\end{mcitethebibliography}


\providecommand{\latin}[1]{#1}
\makeatletter
\providecommand{\doi}
  {\begingroup\let\do\@makeother\dospecials
  \catcode`\{=1 \catcode`\}=2\doi@aux}
\providecommand{\doi@aux}[1]{\endgroup\texttt{#1}}
\makeatother
\providecommand*\mcitethebibliography{\thebibliography}
\csname @ifundefined\endcsname{endmcitethebibliography}
  {\let\endmcitethebibliography\endthebibliography}{}
\begin{mcitethebibliography}{2}
\providecommand*\natexlab[1]{#1}
\providecommand*\mciteSetBstSublistMode[1]{}
\providecommand*\mciteSetBstMaxWidthForm[2]{}
\providecommand*\mciteBstWouldAddEndPuncttrue
  {\def\EndOfBibitem{\unskip.}}
\providecommand*\mciteBstWouldAddEndPunctfalse
  {\let\EndOfBibitem\relax}
\providecommand*\mciteSetBstMidEndSepPunct[3]{}
\providecommand*\mciteSetBstSublistLabelBeginEnd[3]{}
\providecommand*\EndOfBibitem{}
\mciteSetBstSublistMode{f}
\mciteSetBstMaxWidthForm{subitem}{(\alph{mcitesubitemcount})}
\mciteSetBstSublistLabelBeginEnd
  {\mcitemaxwidthsubitemform\space}
  {\relax}
  {\relax}

\bibitem[Weller \latin{et~al.}(2015)Weller, Weber, Henry, Di~Pumpo, and
  Hansen]{Weller:2015jo}
Weller,~M.~T.; Weber,~O.~J.; Henry,~P.~F.; Di~Pumpo,~A.~M.; Hansen,~T.~C.
  {Complete structure and cation orientation in the perovskite photovoltaic
  methylammonium lead iodide between 100 and 352 K }. \emph{Chem. Commun.}
  \textbf{2015}, \emph{51}, 4180--4183\relax
\mciteBstWouldAddEndPuncttrue
\mciteSetBstMidEndSepPunct{\mcitedefaultmidpunct}
{\mcitedefaultendpunct}{\mcitedefaultseppunct}\relax
\EndOfBibitem
\end{mcitethebibliography}

\end{document}


\renewcommand{\figurename}{Figure S}
	\renewcommand{\tablename}{Table S}

\newpage
\listoffigures
\listoftables
\newpage

	\begin{figure}[htp]
		\includegraphics[height=35mm]{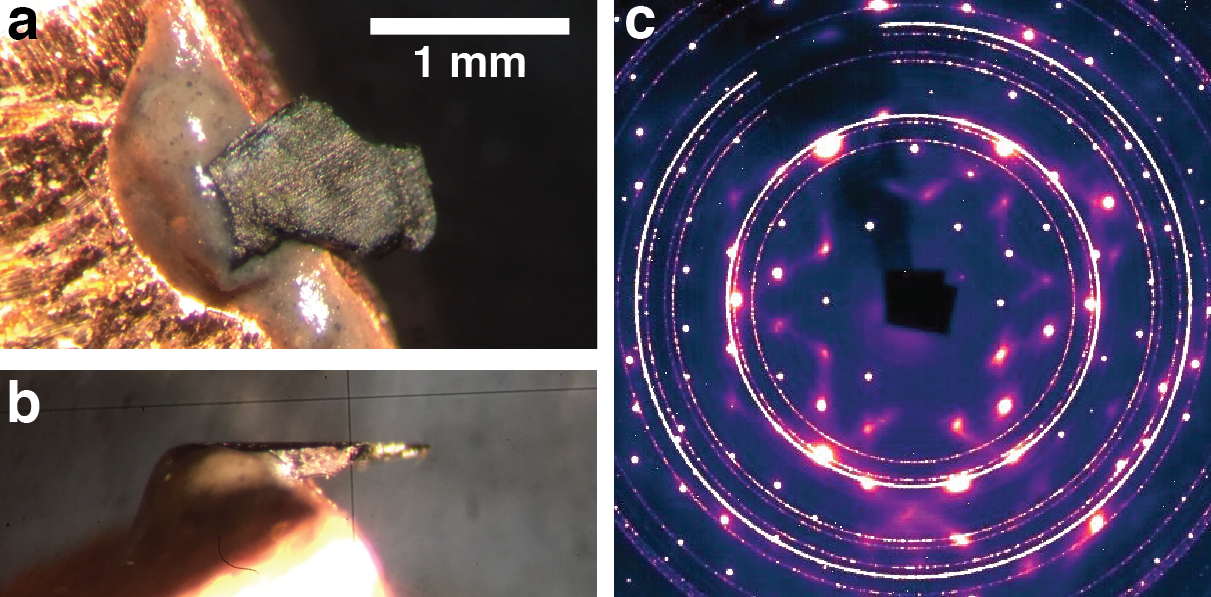}
		\caption[Image of the polished crystal]{Image of polished crystal, as mounted, \textbf{(a)} from top and \textbf{(b)} from side. \textbf{(c)} X-ray scattering images showing [111] zone axis diffraction and diffuse scattering between Bragg peaks. The rings correspond to the diffraction of the beryllium dome which is mostly transparent to X-rays.}
		\label{fig:xtal}
	\end{figure}

\begin{figure}[htp]
	\includegraphics[height=50mm]{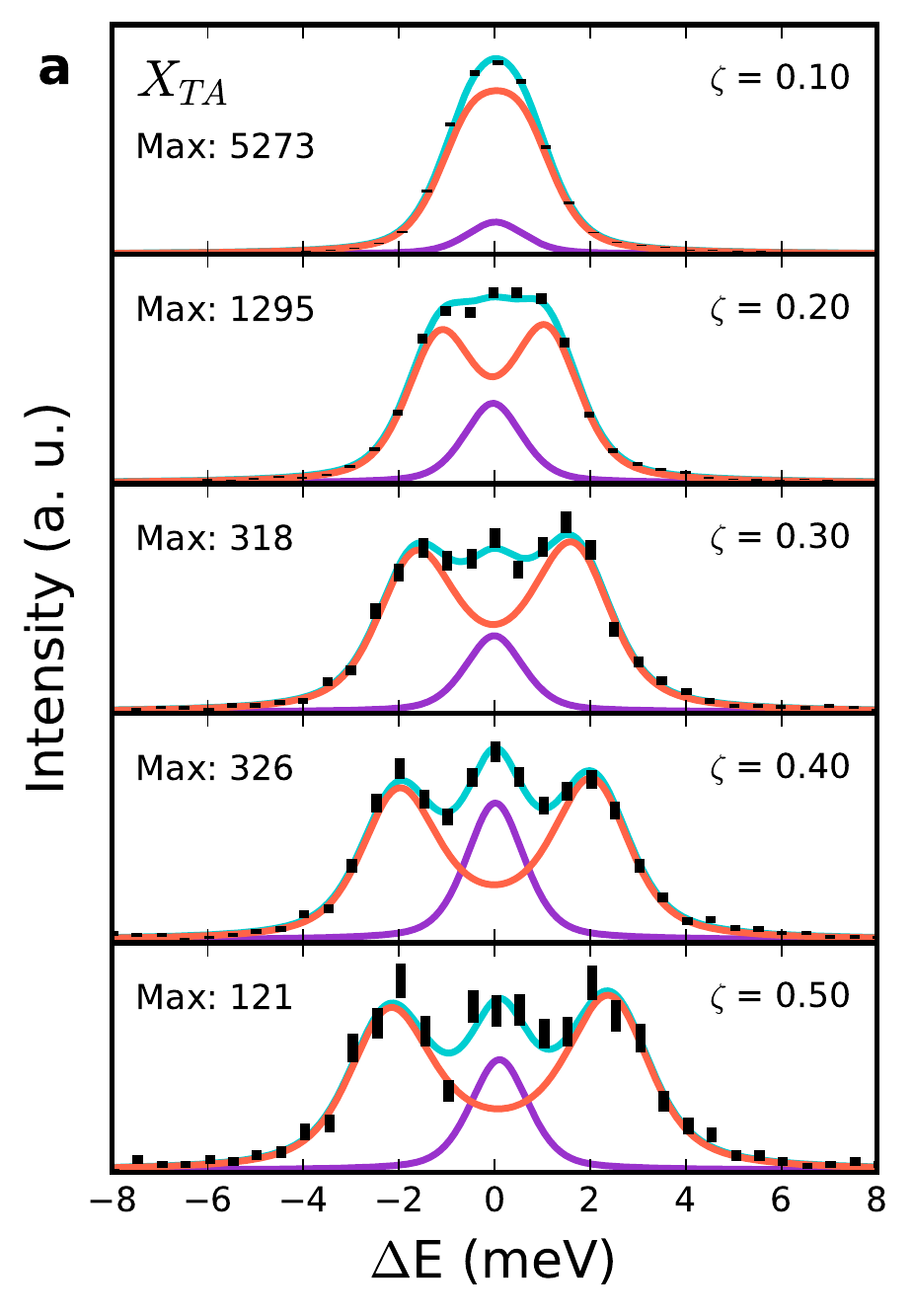}
	\includegraphics[height=50mm]{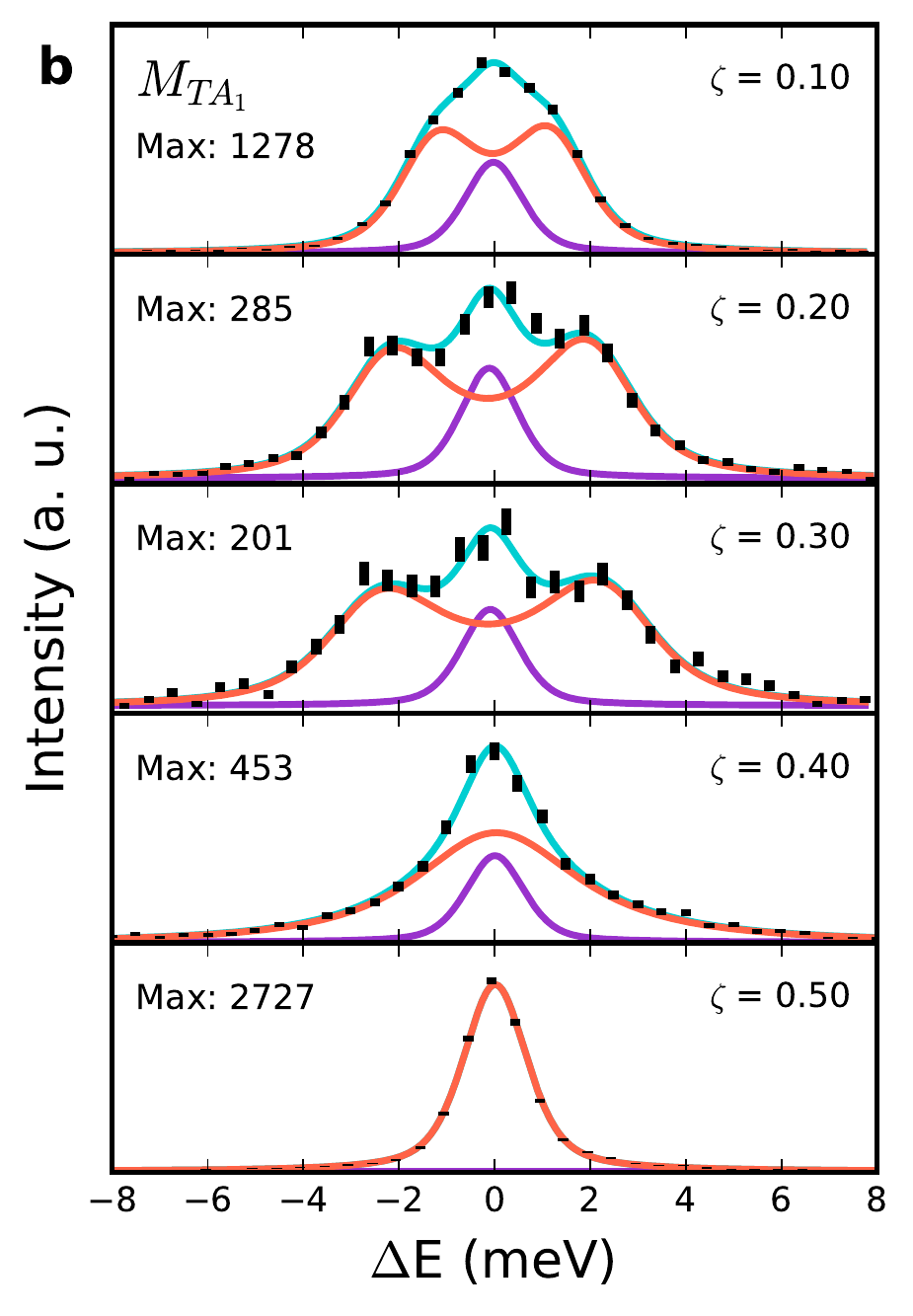}
	\includegraphics[height=50mm]{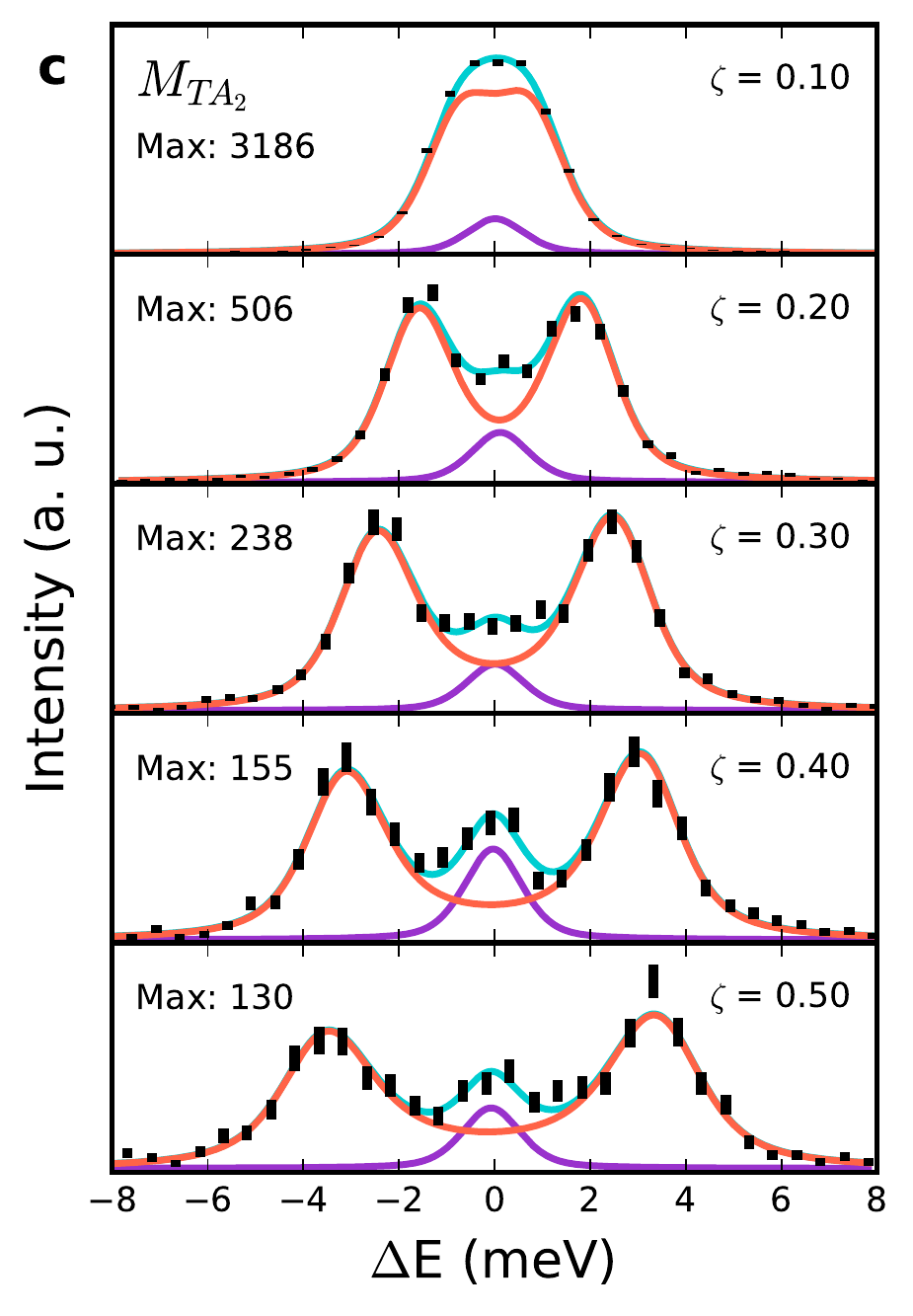}
	\includegraphics[height=50mm]{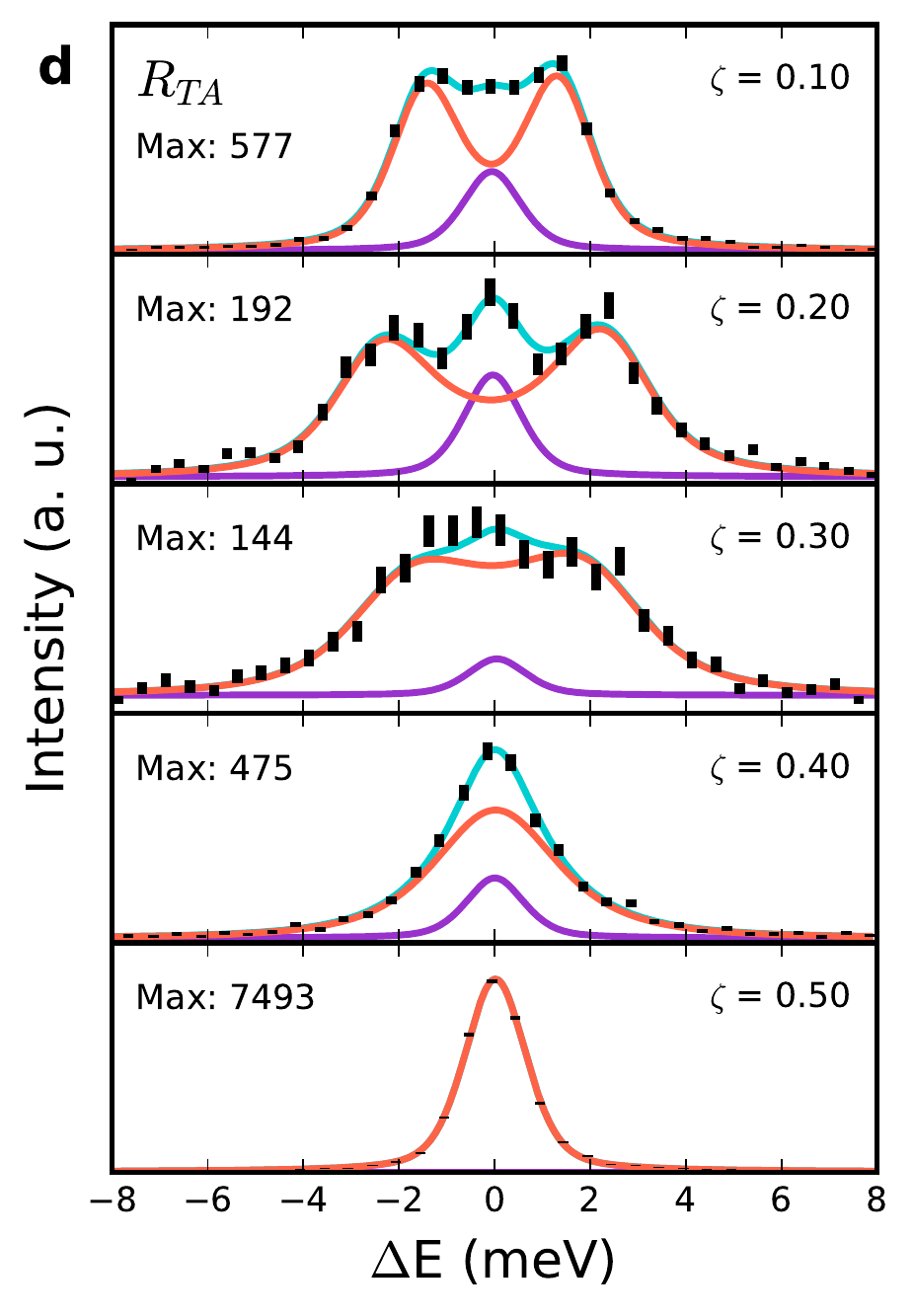}\\
	\includegraphics[height=50mm]{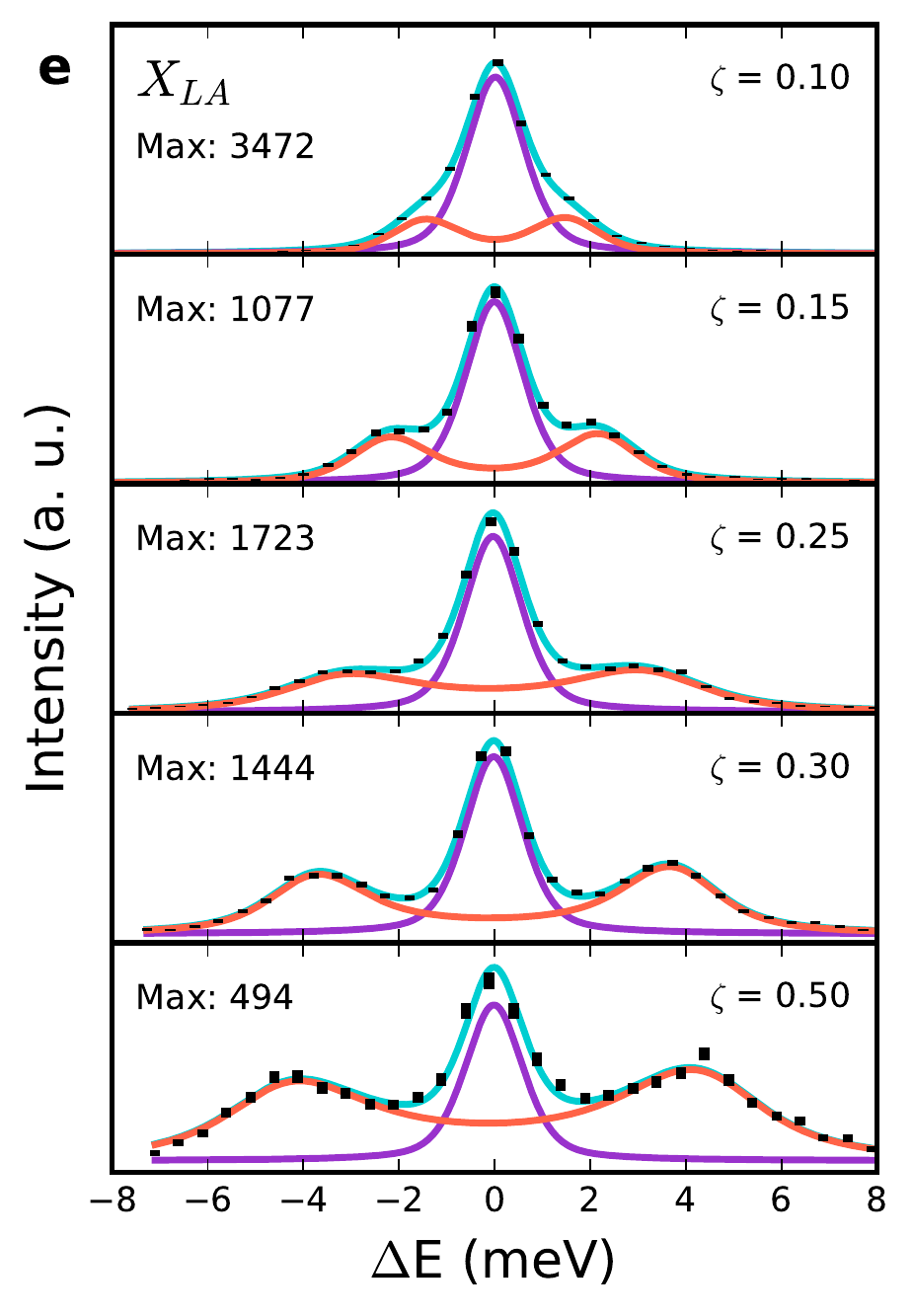}
	\includegraphics[height=50mm]{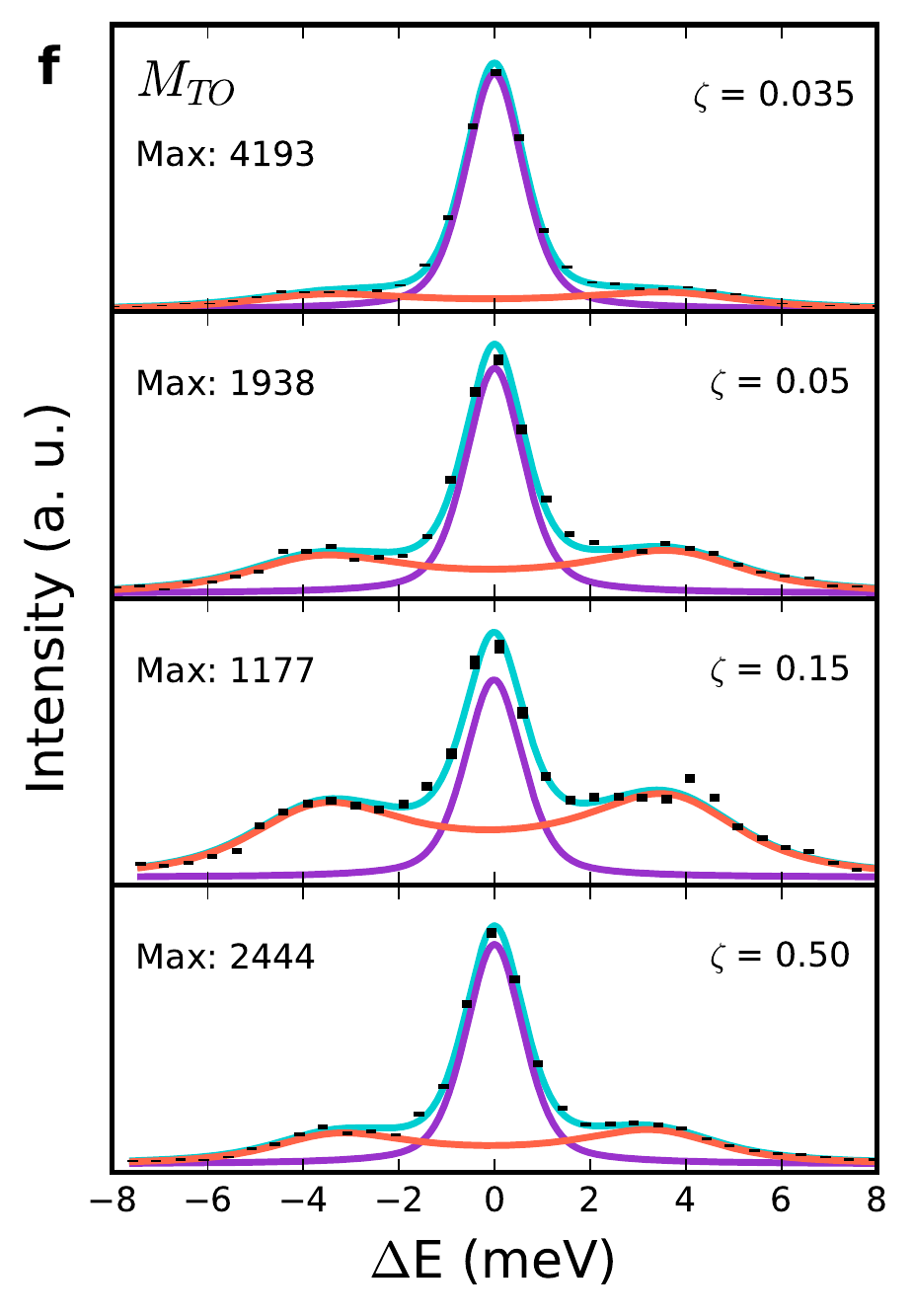}
	\includegraphics[height=50mm]{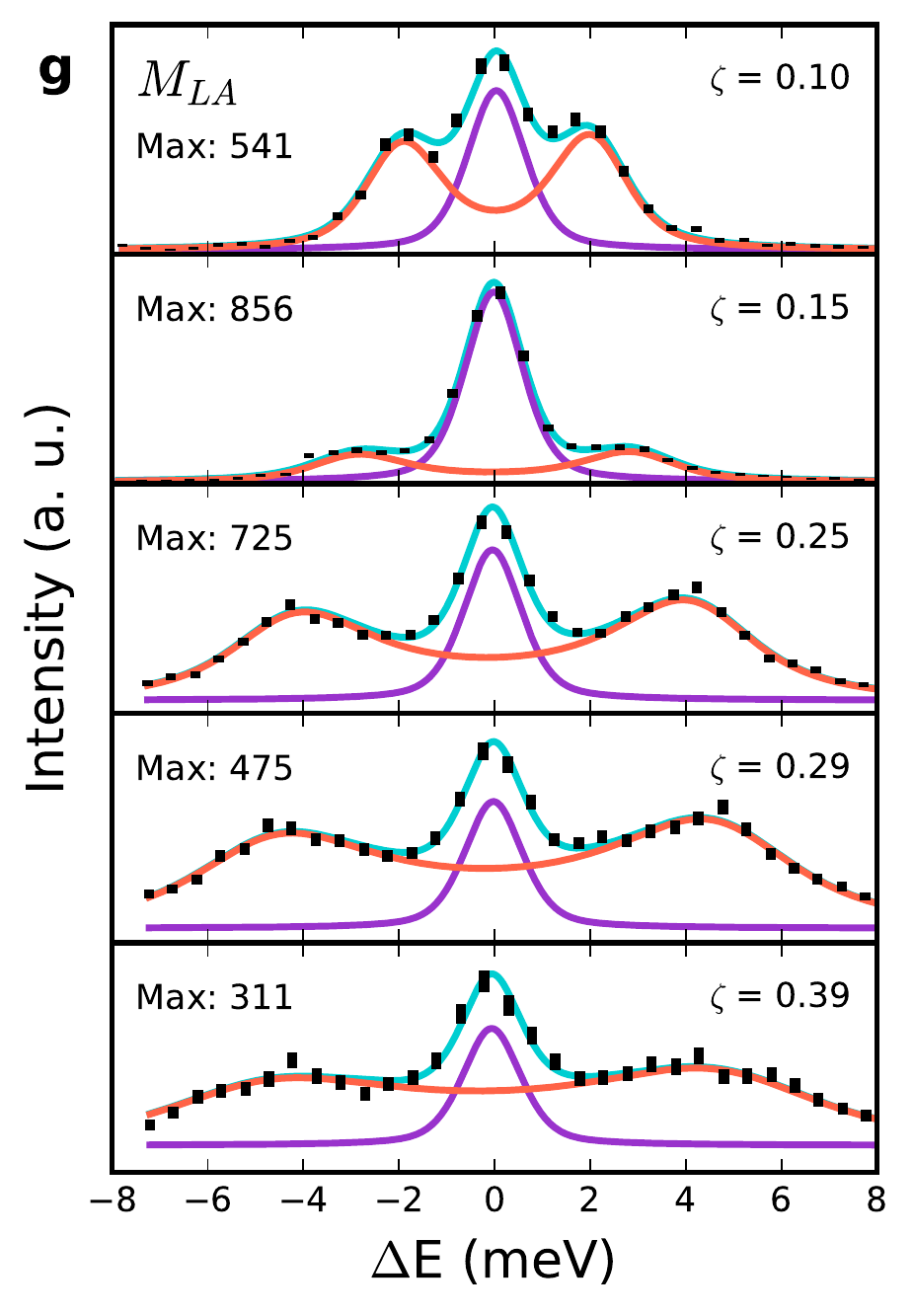}
	\includegraphics[height=50mm]{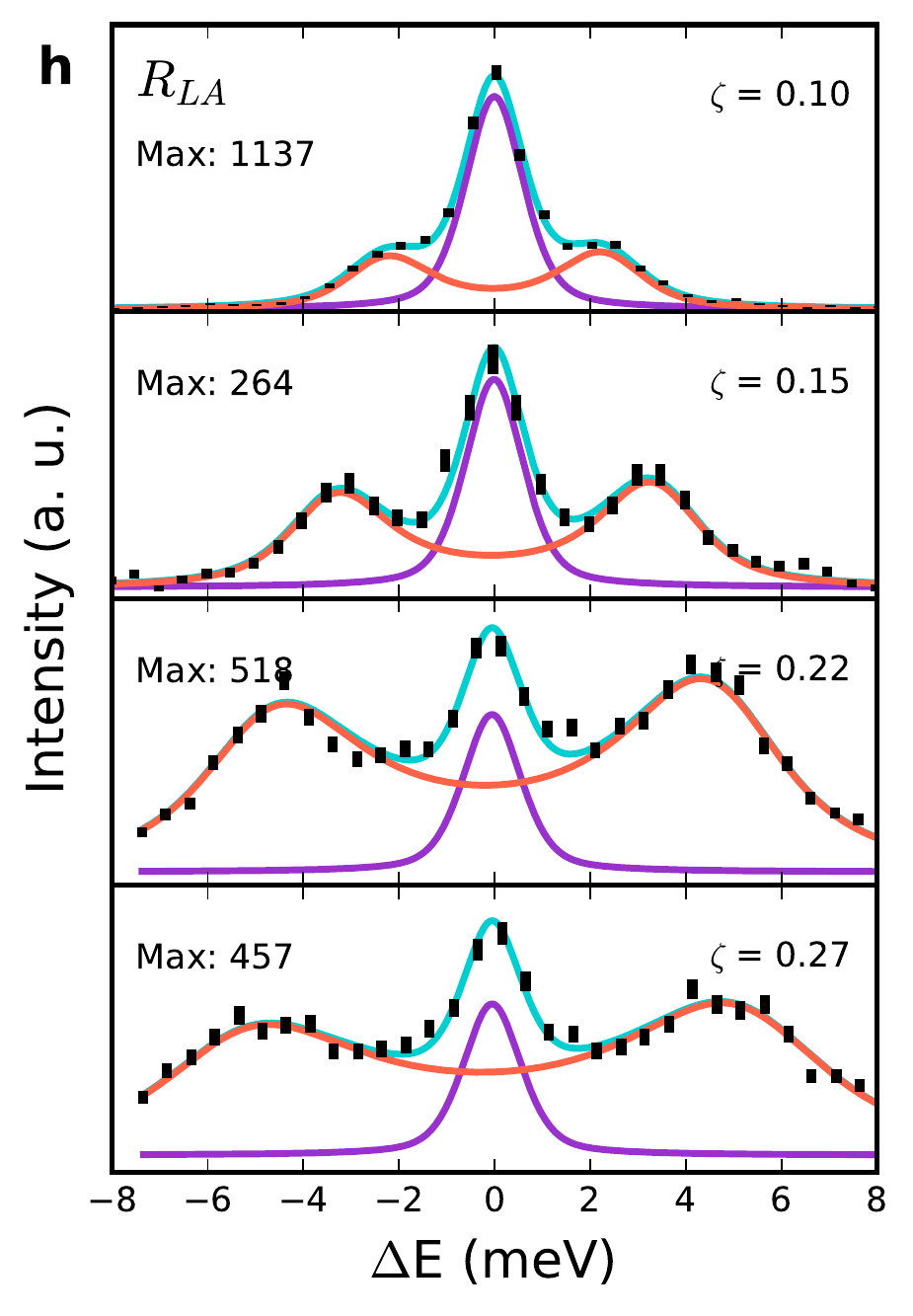}
	\caption[Fitting of the inelastic X-ray scattering spectra]{Fitting of the inelastic X-ray scattering spectra. The central pseudo-Voigt peak, damped harmonic oscillator, and total fit are plotted in purple, orange, and teal respectively.}
	\label{fig:fitting}
\end{figure}

\begin{table}[htp]
\caption{Extracted elastic constants ($C$) and bulk modulus ($K$).}
\begin{center}
\begin{tabular}{ c c c c }
$C_{11}$ [GPa] & $C_{12}$ [GPa] & $C_{44}$ [GPa] & $K$ [GPa]\\
\hline
$25 \pm 2$ & $7 \pm 2$ & $4 \pm 1$ & $13 \pm 2$
\end{tabular}
\end{center}
\label{tab:elastics}
\end{table}%

\begin{figure}
	\includegraphics[width=70mm]{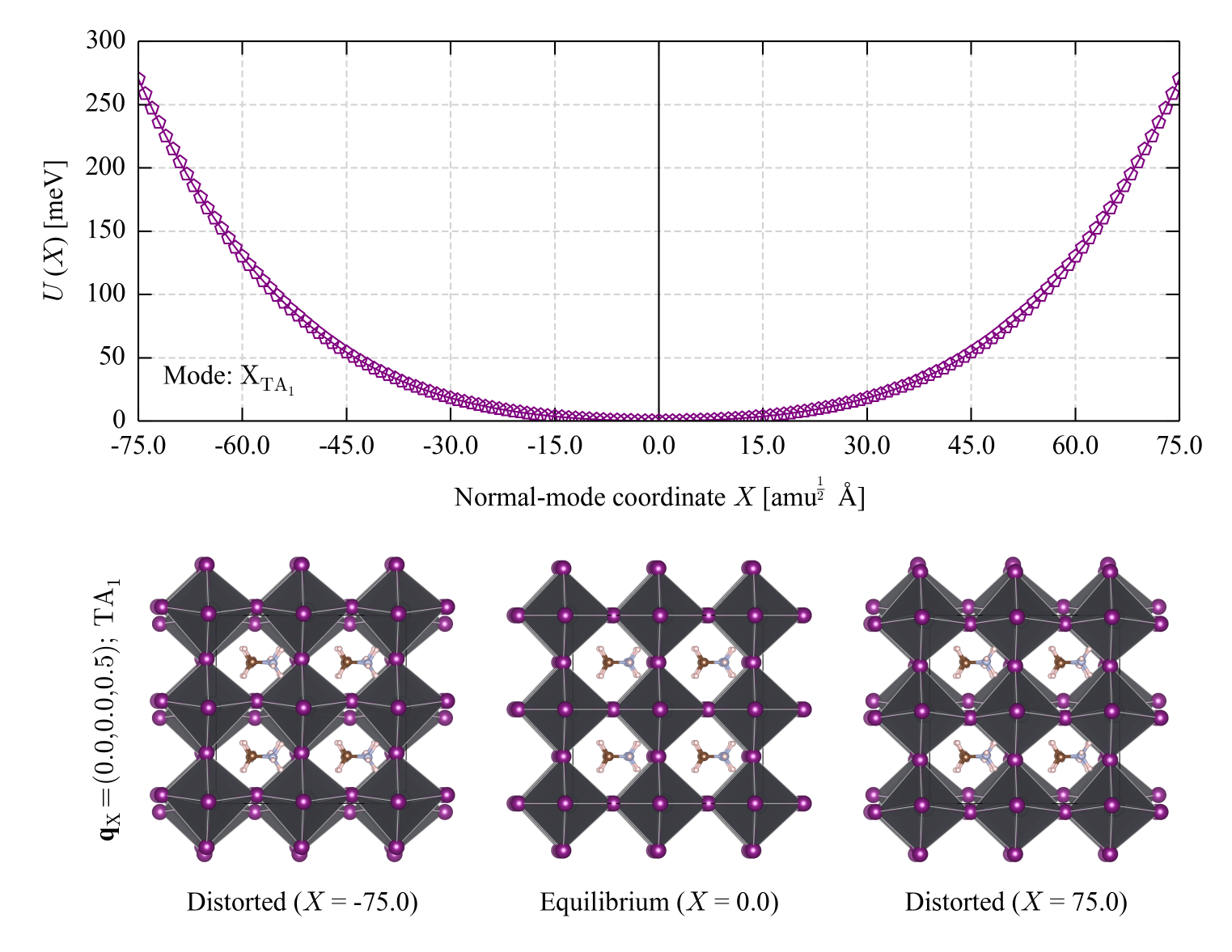}\\
	\includegraphics[width=70mm]{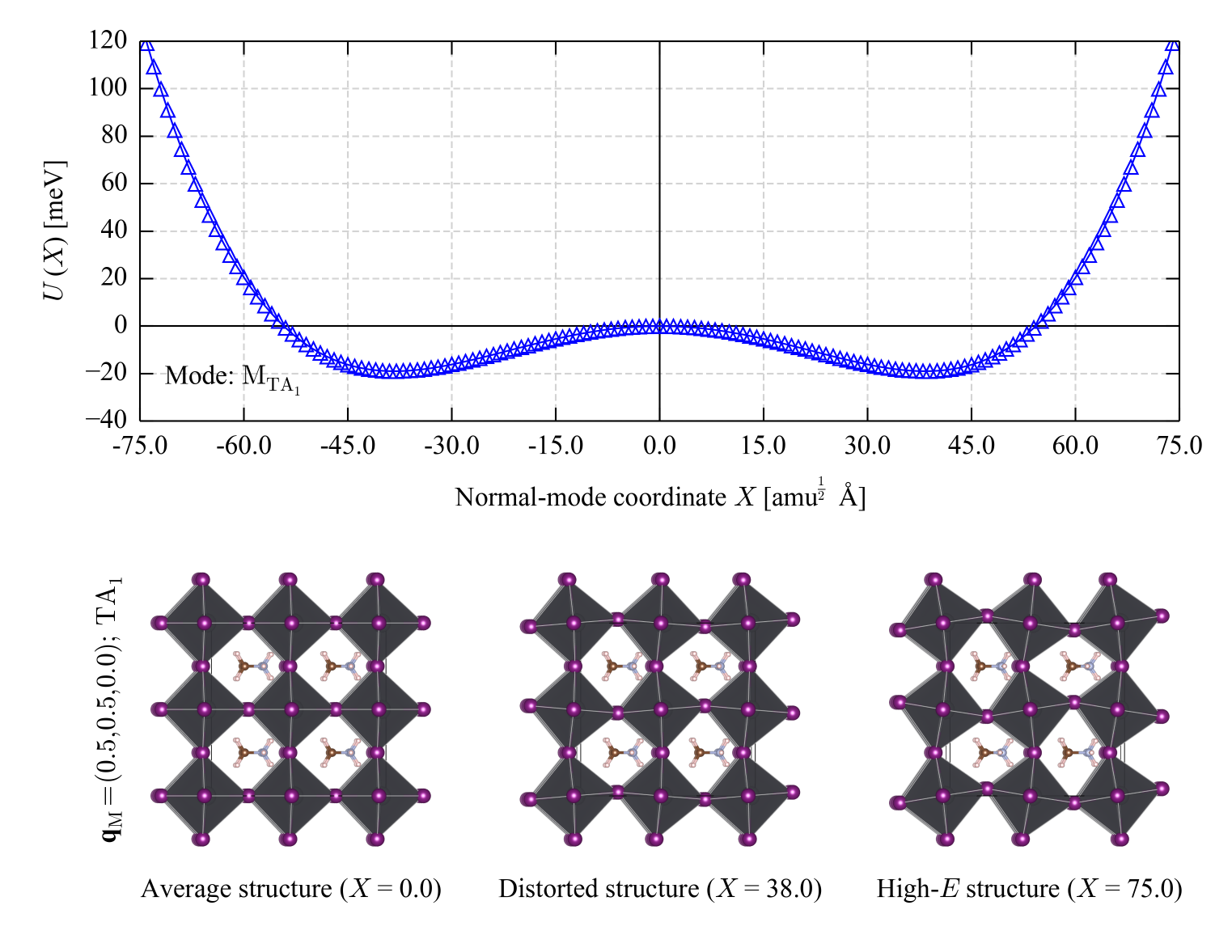}\\ 
	\includegraphics[width=70mm]{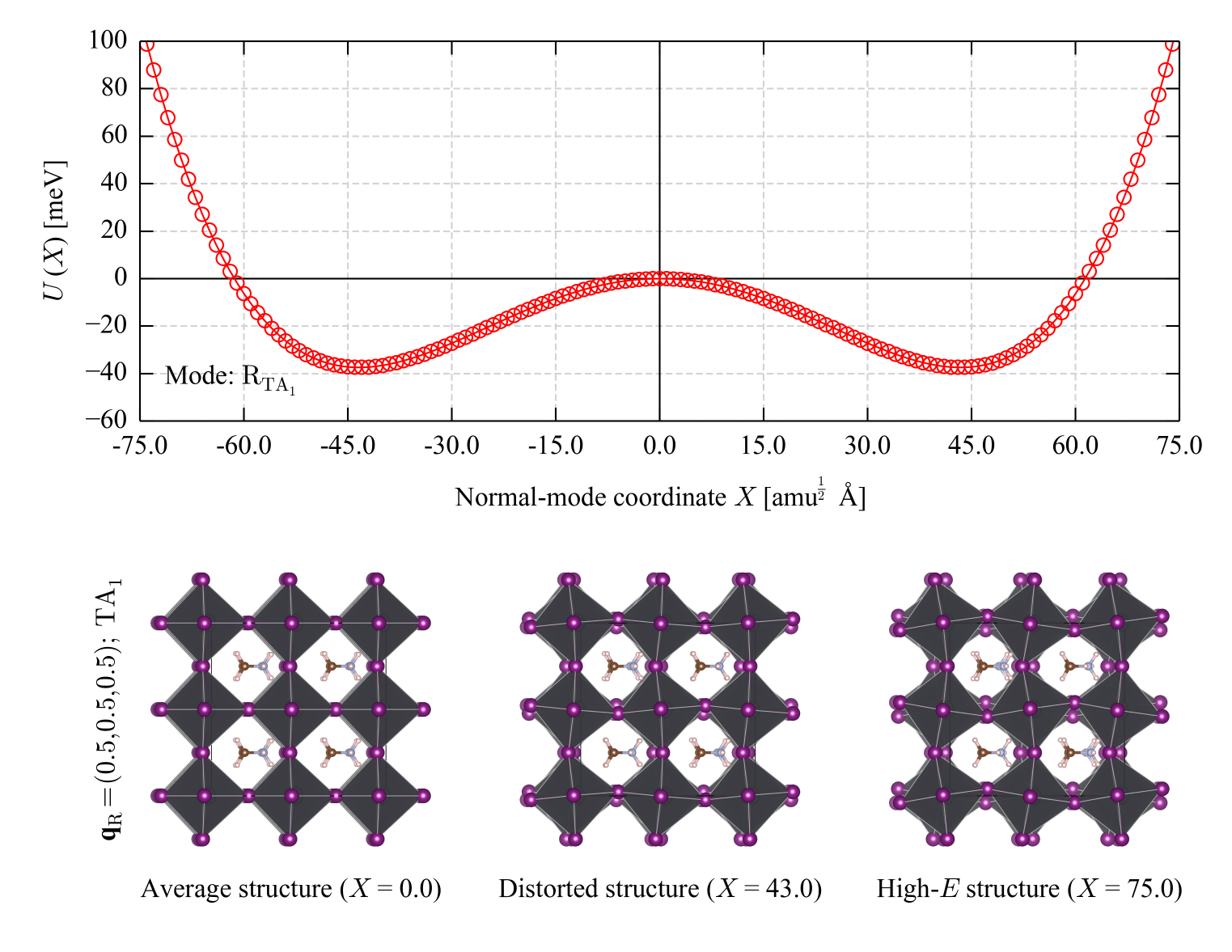}\\
	\includegraphics[width=70mm]{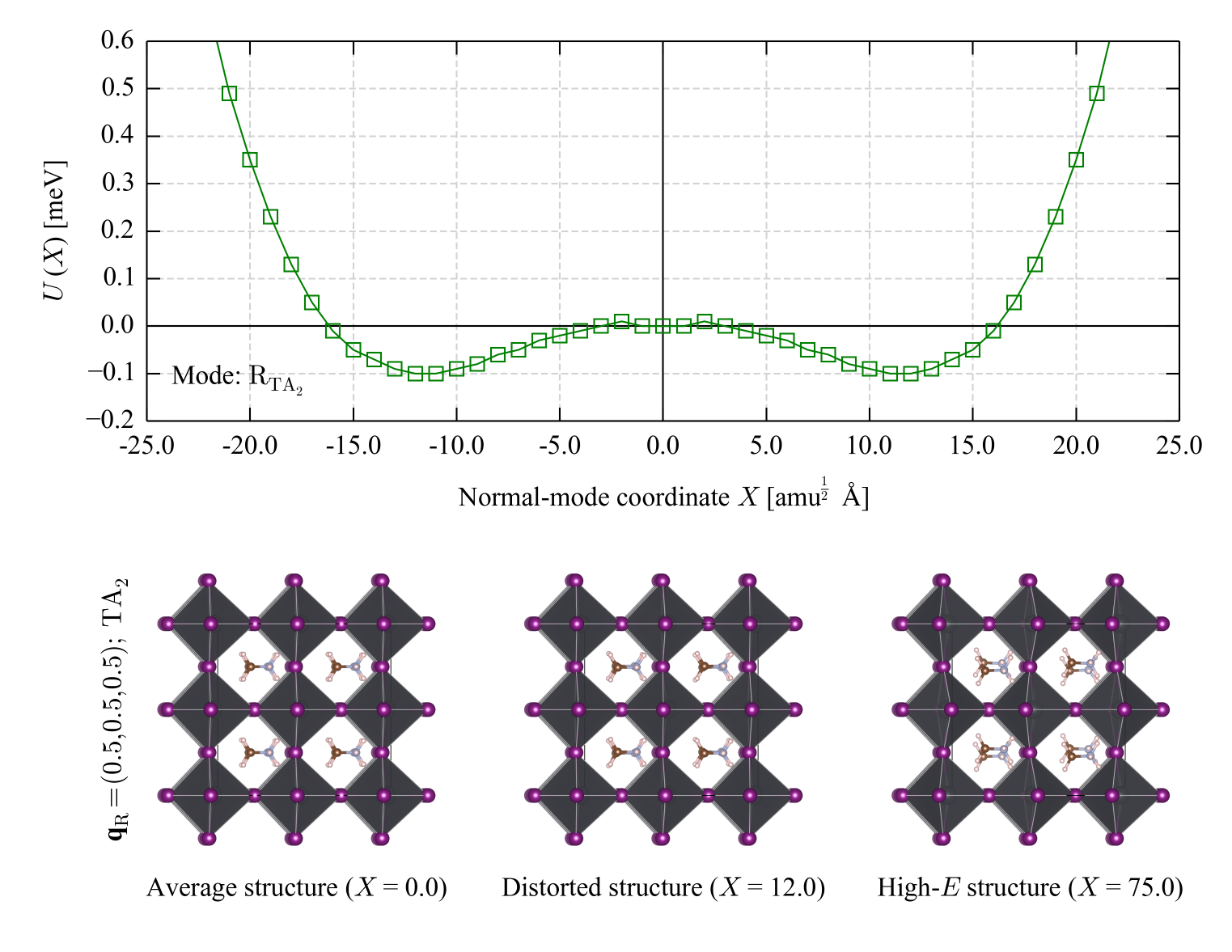}
	\caption[Potential energy surfaces calculated for ion displacement in \MAPI]{Potential energy surfaces calculated for ion displacement along the imaginary phonon eigenvectors at the $X_{TA}$, $M_{TA_1}$, and $R_{TA}$ zone-edge points. Surfaces for inequivalent $R$ zone-edge points are drawn ($R_{TA_1}$ and $R_{TA_2}$). Surfaces for the soft modes show two energy minima displaced from 0.}
	\label{fig:potentials}
\end{figure}

\begin{figure}
	\includegraphics[width=70mm]{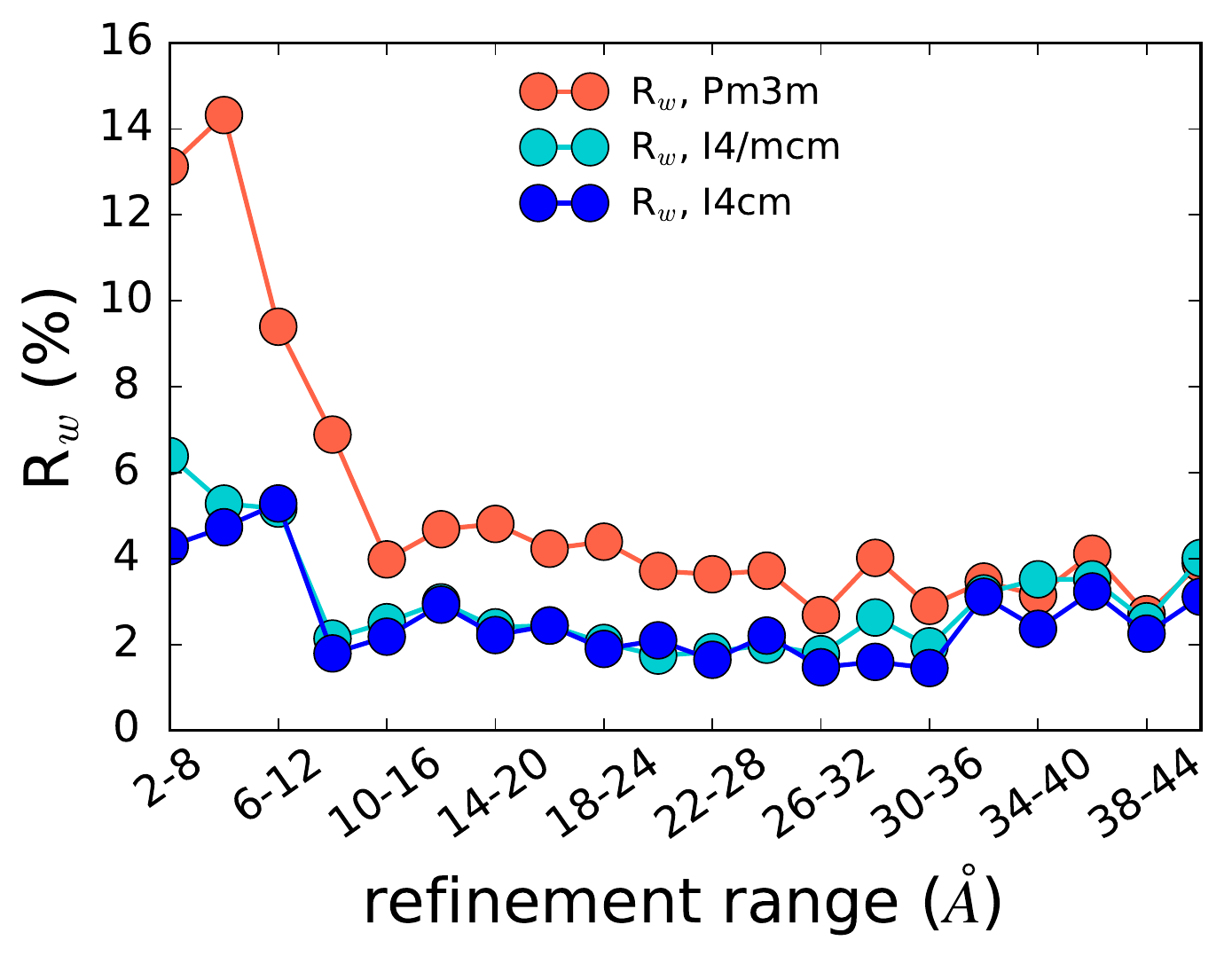}
	\caption[Comparison of PDF refinements across different refinement ranges]{Comparison of refinements across different refinement ranges of three structural models (cubic $Pm3m$, centrosymmetric tetragonal $I4/mcm$, and non-centrosymmetric tetragonal $I4cm$) to the experimental PDF of \MAPI\ at 350~K. $R_w$ is the quality of the fit, lower is better. 
	At low refinement ranges, the structure is best modeled as tetragonal, but as the refinement range increases to higher $r$, the cubic model has excellent agreement with the data.
		The $Pm3m$ and $I4/mcm$ models are based upon the powder neutron diffraction structures reported by Weller and coworkers~\cite{Weller:2015jo}. $I4cm$ structure was derived from the $I4/mcm$ structure by letting the Pb atoms freely displace along the c-direction of the unit cell during the refinement in the 2-8~$\text{\AA}$ range.}
	\label{fig:refranges}
\end{figure}

\begin{figure}
	\includegraphics[width=100mm]{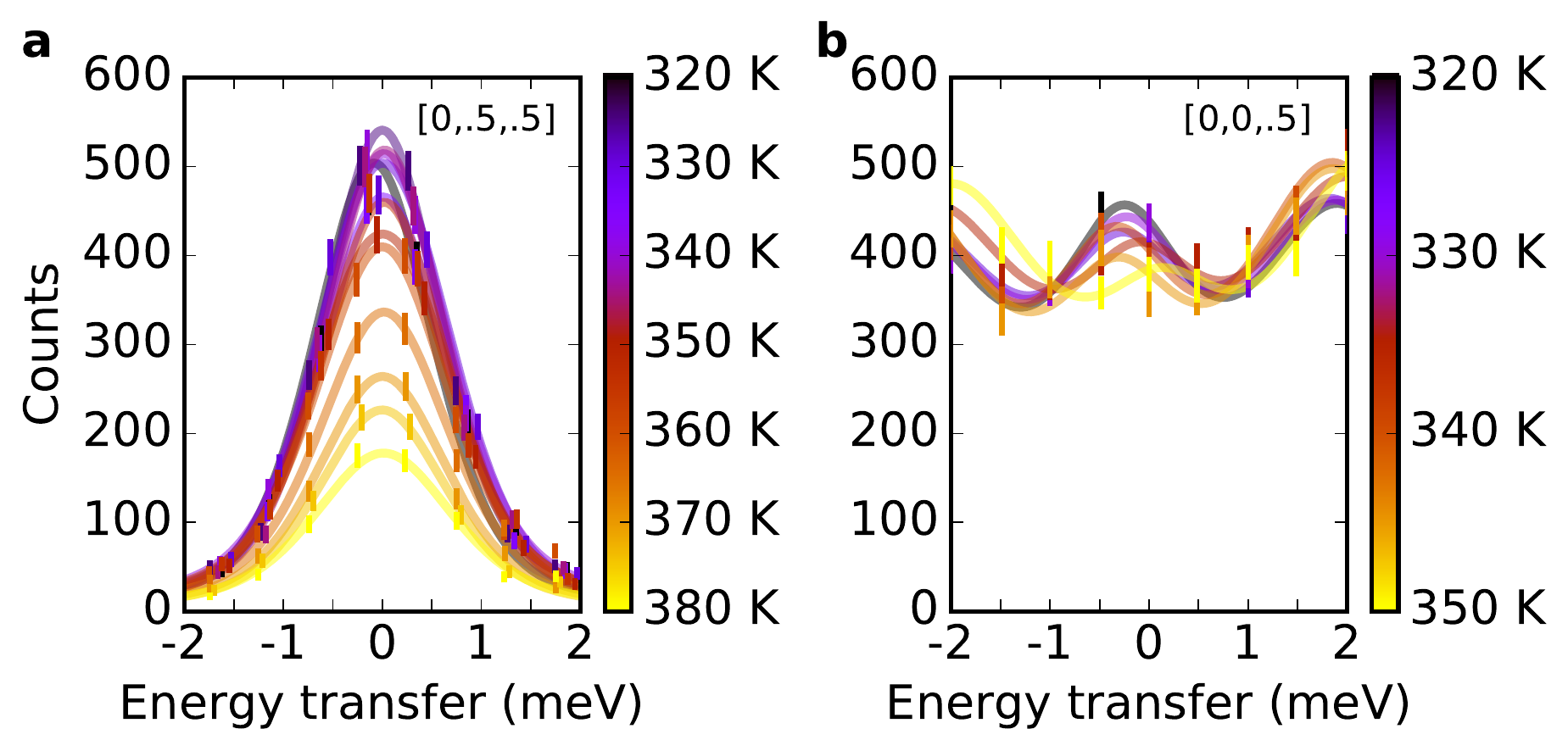}
	\caption[Energy scans as a function of temperature]{Energy scans at the \textbf{(a)} $M$- and \textbf{(b)} $X$-points as a function of temperature. While the energy scans at the $M$-point show a clear temperature dependence, intensity at the $X$-point declines slightly with decreasing temperature if it changes at all, which might be related to changes in the thermal population.}
	\label{fig:TdepXM}
\end{figure}

\begin{figure}
	\includegraphics[width=62mm]{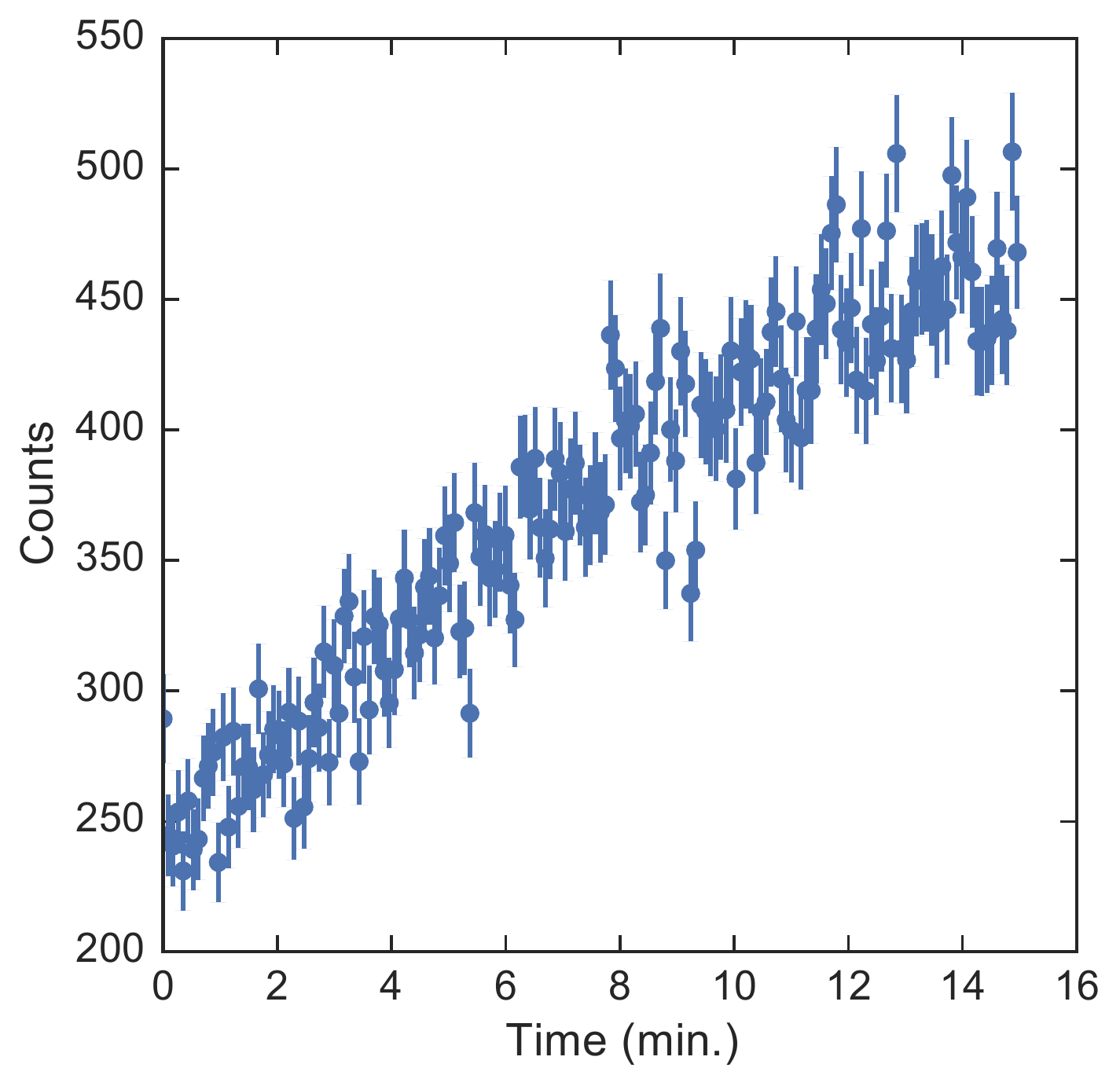}
	\caption[Sample instability to beam exposure]{Sample instability. Over time, we observe a slow increase in the central quasi-elastic peak ($\hbar\omega=0$), regardless of the reciprocal lattice coordinate of the measurement. This appears consistent with a moderate increase of the disorder of the crystal during beam irradiation. We observe no change in quasi-elastic scattering intensity if the shutter is kept closed in between measurements. This data was collected at $[0.0\ 3.5\ 4.5]$ at 350 K.}
	\label{fig:degradation}
\end{figure}
\clearpage{}
\newpage{}
\bibliography{taviRefs,otherRefs}